\documentclass[pra,aps,twocolumn,longbibliography]{revtex4-1}

\usepackage{graphicx}
\usepackage{bm}
\usepackage{amsmath}
\usepackage{amssymb}
\usepackage{amsthm}
\usepackage{amsfonts}
\usepackage{enumerate}
\usepackage{latexsym}
\usepackage{hyperref}

\begin{document}

\title{Quantum corrections to a spin-orbit coupled Bose-Einstein Condensate}

\author{Long Liang$^{1,2}$ and P\"aivi T\"orm\"a$^{1}$}

\affiliation{
	1, Department of Applied Physics, Aalto University School of Science, FI-00076 Aalto, Finland \\
	2, Computational Physics Laboratory, Physics Unit, Faculty of Engineering and
Natural Sciences, Tampere University, P.O. Box 692, FI-33014 Tampere,
Finland
}

\begin{abstract}
We study systematically the quantum corrections to a  weakly interacting Bose-Einstein condensate with spin-orbit coupling.
We show that quantum fluctuations, enhanced by the spin-orbit coupling, modify quantitatively the mean-field properties such as the superfluid density, spin polarizability, and sound velocity. 
We find that the phase boundary between the plane wave and zero momentum phases is shifted to a smaller transverse field. 
We also calculate the Beliaev and Landau damping rates and find that the Landau process dominates the quasiparticle decay even at low temperature. 
\end{abstract}

\maketitle

\section{Introduction}

The spin-orbit coupling, arising due to the interaction of a particle's spin with its motion in an electric field plays a crucial role in various  branches of physics, including topological insulators \cite{RevModPhys.82.3045,RevModPhys.83.1057}, topological semimetals \cite{doi:10.1146/annurev-conmatphys-031016-025458,RevModPhys.90.015001}, and Majorana fermions \cite{RevModPhys.87.137}.
In bosonic systems, the interplay of the interparticle interaction and spin-orbit coupling gives rise to exotic Bose-Einstein condensates which have been investigated in  a rich variety of  systems, including magnons \cite{Ruegg2003,2004EL.....68..275S,Demokritov2006}, excitons \cite{PhysRevLett.98.166405,PhysRevLett.103.086404,2012Natur.483..584H,PhysRevLett.110.246403}, exciton-polaritons \cite{RevModPhys.85.299,Byrnes2014,PhysRevX.5.011034,PhysRevLett.120.097401,doi:10.1063/1.4995385,doi:10.1021/acsphotonics.8b00536},  
and ultracold atoms \cite{2011Natur.471...83L, PhysRevLett.109.115301, PhysRevLett.114.105301, 2016Sci...354...83W}.

For a weakly interacting Bose-Einstein condensate, the mean-field theory provides a reliable description of various physical properties \cite{RevModPhys.71.463}. 
To reveal beyond mean-field effects, one method is to reach the strongly interacting regime, which can be achieved in exciton-polaritons because of the strong coupling between the exciton and photon \cite{RevModPhys.85.299,Kasprzak2006,Balili1007,PhysRevLett.118.247402,Fink2018},
and for ultracold atoms strong interactions are accessible by  means of Feshbach resonances \cite{PhysRevLett.101.135301,PhysRevLett.102.090402,PhysRevLett.107.135301}. However, strong interactions reduce the lifetime of Bose-Einstein condensates significantly.
Another method is to fine tune the interaction parameters such that the mean-field interactions almost cancel out \cite{PhysRevLett.115.155302,2017NJPh...19k3043L,Cabrera301, PhysRevLett.120.235301,PhysRevLett.121.173403}, making the quantum fluctuations unmasked. 
 The spin-orbit coupling provides an alternative way to enhance interaction effects due to the increased density of states \cite{2015RPPh...78b6001Z}. 
However, only a handful of theoretical studies have addressed the beyond mean-field effects \cite{PhysRevLett.109.025301,CuiZhouPRA2013,0953-4075-46-13-134007,PhysRevA.95.051601, Beliaev_SOC},  
and a thorough analysis of the quantum fluctuations in spin-orbit coupled bosonic systems is still lacking.

In this paper, we systematically investigate the quantum corrections to a spin-orbit coupled Bose-Einstein condensate. 
We study a model system that is simple and general, potentially realizable in various platforms and already implemented with ultracold atom experiments \cite{2011Natur.471...83L, PhysRevLett.109.115301, PhysRevLett.114.105301}.
The model shows three novel condensation phases \cite{2011Natur.471...83L,SOC_BosonsStringari2012a}, namely the stripe, plane wave, and zero momentum phases. 
To demonstrate the interplay between interaction and spin-orbit coupling, we focus on the zero momentum phase, which is the simplest case capturing the essential physics of spin-orbit coupling and interactions.  

We calculate quantum corrections to a number of physical properties, including the superfluid density, spin polarizability, and sound velocity. 
The superfluid density at the mean-field phase transition point between the plane wave and zero momentum phases becomes nonzero due to quantum fluctuations, and  as a result, the phase transition point is shifted towards a smaller transverse field. 
The spin polarizability diverges at the corrected phase transition point but remains finite at the mean-field phase boundary, which seems to be consistent with a recent experiment \cite{PhysRevLett.109.115301}.
The sound velocity also acquires quantitative corrections, which may be detected in current ultracold atom experiments and provides a way to explore the beyond mean-field effects.
Finally, we obtain an analytical result for the Landau decay rate of phonons at low temperature. 
Unlike the Beliaev decay predicted in \cite{Beliaev_SOC}, the Landau damping  is not suppressed in the direction of spin-orbit coupling, making it the dominant mechanism for the quasiparticle decay. 

\section{ The model system}

We consider a generic model of a spin-1/2 Bose gas with spin-orbit coupling, described by the single particle Hamiltonian (we set $\hbar=m=1$)
\begin{eqnarray}
h_0=\frac{(p_x-k_0\sigma_z)^2+p^2_y+p^2_z}{2}+\frac{\Omega}{2}\sigma_x,
\end{eqnarray}
where $\sigma_i$ with $i=x,y,z$ are the $2\times 2$ Pauli matrices.
The one dimensional spin-orbit coupling, characterized by $k_0$, appears in 
many realistic systems, including  ultracold gases \cite{2011Natur.471...83L, PhysRevLett.109.115301, PhysRevLett.114.105301,PhysRevLett.109.095301, PhysRevLett.109.095302} and semiconducting nanowires \cite{Quay2010,Mourik1003,Das2012}. 
The model applies to several systems but to compare with experiments, we consider the cold atom setup where $k_0$ is given by the momentum transfer from the two Raman laser beams and $\Omega$ is the Rabi frequency of the Raman beams. 
The interaction between the particles can be written as
\begin{eqnarray}
H_{\mathrm{int}}= 
\frac{1}{2}\sum_{\sigma\sigma'}\int \mathrm{d}^3\mathbf{r}~ g_{\sigma\sigma'}n_\sigma(\mathbf{r})  n_{\sigma'}(\mathbf{r}),
\end{eqnarray}
where $n_{\sigma}$ is the density of particles with spin $\sigma=\uparrow, \downarrow$, and $g_{\sigma\sigma'}=4\pi a_{\sigma\sigma'}$ are the interaction strengths in different spin channels, with $a_{\sigma\sigma'}$ being the corresponding $s$-wave scattering lengths in case of ultracold quantum gases. 
In the following we assume $g_{\uparrow\uparrow}=g_{\downarrow\downarrow}\equiv g$ and $g_{\uparrow\downarrow}\equiv g'$, and correspondingly, $a_{\uparrow\uparrow}=a_{\downarrow\downarrow}\equiv  a$ and $a_{\uparrow\downarrow}\equiv a'$. 
It is convenient to define interaction parameters $G_1=g_+\rho$ and $G_2=g_-\rho$ with $g_{\pm}=(g\pm g')/2$ and $\rho=N/V$ being the total particle density. 
In  recent experiments \cite{2011Natur.471...83L,PhysRevLett.109.115301,PhysRevLett.114.105301}, $^{87}$Rb atoms are employed and the interaction is almost $SU(2)$ invariant, with $G_2/G_1\approx 10^{-3}$. 
The typical interaction parameter is
$G_1\approx 0.24k^2_0$ with the peak density $\rho\approx0.57k^3_0$ \cite{PhysRevLett.109.115301}. 
The dimensionless parameter $\sqrt{a^3 \rho}\approx 0.004$ is small, ensuring that the condensate is in the weakly interacting regime and the perturbation calculations are controlled.

The mean-field phase diagram of this model has been extensively investigated, for a review see \cite{2015RPPh...78b6001Z}.
For small Rabi frequency, the condensate wave function is a superposition of two plane waves with different momenta, characterizing the stripe phase with density modulations in the ground state. 
In this phase, both the translational and $U(1)$ symmetries are broken, and therefore there are two branches of gapless excitations. 
Increasing the Rabi frequency $\Omega$, the system enters the plane wave phase, in which the bosons condense in a single plane wave state. 
There is only one branch of gapless excitations in this phase and the energy dispersion contains a roton minimum at finite momentum. 
Further increasing the Rabi frequency such that $\Omega>\Omega_{c,\mathrm{mf}}\equiv 2k^2_0-2G_2$, the system enters the zero momentum phase, where the roton minimum disappears and the phonon excitation spectrum resembles that of a Bose-Einstein condensate without spin-orbit coupling.
To reveal the essential effect of interactions, we focus on the simplest zero momentum phase to reduce the effect of nontrivial mean-field energy dispersions in the plane wave and stripe phases.

The ground state wave function for the zero momentum phase is described by a spinor $\psi=\sqrt{\rho/2}[1,-1]^T$. 
To characterize excitations on top of the condensate, we introduce phase and number fluctuations, and write the spinor field as
\begin{eqnarray}\label{Eq:field_parameterization2}
\psi=\frac{e^{i \phi}}{\sqrt{2}}\left[\begin{array}{c}
\sqrt{\rho+\zeta_1} e^{i\varphi}\\ 
- \sqrt{\rho+\zeta_2} e^{-i\varphi}
\end{array} \right],
\end{eqnarray} 
where $\phi$ is the total and $\varphi$ is the relative phase fluctuations of the condensate, and $\zeta_1$ and $\zeta_2$ are the density fluctuations for spin up and spin down particles, respectively. 

\begin{figure} 
	\includegraphics[width=\columnwidth]{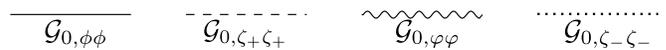}
    \caption{Feynman digrams for diagonal elements of the mean-field Green's function.}\label{Fig:G0}
\end{figure}

We use the imaginary time path integral formalism. 
The Lagrangian density is obtained through the Hamiltonian as 
\begin{eqnarray}
\mathcal{L}&=&\psi^\dagger(\partial_\tau+h_0-\mu)\psi
 + \frac{g_+}{2}(\psi^\dagger\psi)^2+\frac{g_-}{2}(\psi^\dagger\sigma_z\psi)^2,~
\end{eqnarray}
with $\mu$ being the chemical potential. 
It is convenient to introduce the density and spin fluctuations, $\zeta_+=(\zeta_1+\zeta_2)/2$ and $\zeta_-=(\zeta_1-\zeta_2)/2$, which are  conjugate to $\phi$ and $\varphi$, respectively. 
We then expand  $\mathcal{L}$ in terms of the new variables, and up to the second order,  we get the mean-field Lagrangian density
\begin{eqnarray}
\mathcal{L}_{\mathrm{mf}}&=&\left(G_1+\frac{k^2_0-\Omega}{2}-\mu\right)\zeta_+ -\mu\rho+
\frac{g_+\rho^2}{2}\nonumber\\
&&
+\frac{1}{2} [\phi,\zeta_+,\varphi,\zeta_-]\mathcal{G}^{-1}_{0}[\phi,\zeta_+,\varphi,\zeta_-]^{T},
\end{eqnarray}
where the mean-field Green's function in the momentum and frequency representation is 
\begin{eqnarray}
  \mathcal{G}^{-1}_{0}(i\omega_n,\mathbf{q})=\left[\begin{array}{cccc}
   \mathcal{A}(\mathbf{q}) & -\omega_n & 0 & -i k_0 q_x \\ 
   \omega_n & \mathcal{B}(\mathbf{q}) & i k_0 q_x & 0\\ 
  0 & -i k_0 q_x & \mathcal{C}(\mathbf{q}) & -\omega_n  \\ 
 i k_0 q_x   & 0 & \omega_n & \mathcal{D}(\mathbf{q})
   \end{array} 
   \right],
  \end{eqnarray}  
here $\omega_n=2\pi n T$ is the Matsubara frequency (we set $k_B=1$), $\mathcal{A}(\mathbf{q})=\rho \mathbf{q}^2$, $\mathcal{B}(\mathbf{q})=\mathbf{q}^2/(4\rho)+g_+$, $\mathcal{C}(\mathbf{q})=\rho \mathbf{q}^2+2\Omega\rho$, and $\mathcal{D}(\mathbf{q})=\mathbf{q}^2/(4\rho)+g_- +\Omega/(2\rho)$. 
The chemical potential is determined by requiring $\langle \zeta_+\rangle=0$, and at the mean-field level we find  $\mu=G_1+(k^2_0-\Omega)/2$, so the first order term of $\zeta_+$ vanishes and the mean-field Lagrangian density is quadratic. The diagonal elements of $\mathcal{G}_0$ are represented by Feynman diagrams shown in Fig.~\ref{Fig:G0}.  

\section{ Mean-field results}

Before studying the beyond mean-field corrections, we first present the mean-field predictions of the physical properties we are interested in.
These results are readily obtained from the mean-field Green's function. 

The mean-field excitation energy is determined by $\det \mathcal{G}^{-1}_{0}=0$. In the low momentum limit, we find the gapless phonon dispersion to be
\begin{eqnarray}
\varepsilon_{\mathrm{ph}}= c_0\sqrt{q^2-\frac{2k^2_0 q^2_x}{\Omega+2G_2}}=c_0 q\sqrt{1-\frac{2k^2_0 \cos^2{\theta}}{\Omega+2G_2}}
\equiv c_{\theta}q,~~~\label{Eq:phonon_dispersion}
\end{eqnarray}
where $c_0=\sqrt{G_1}$ and $c_{\theta}$ is the sound velocity which depends on the angle $\theta$  between the directions of the momentum $\mathbf{q}$ and the $x$ axis. 
The mean-field sound velocities $c_y$ and $c_z$ are the same as the usual Bogoliubov sound velocity $c_0$. 
An intriguing feature is  that the mean-field sound velocity in the $x$ direction, $c_x$, vanishes at the phase transition point between the plane wave and zero momentum phases. 
Besides the gapless phononic mode, there also exists a gapped  mode which is dominated by spin excitation, with the mean-field gap given by $\Delta_0=\sqrt{\Omega (\Omega+2G_2)}$. 
 
The density and spin response functions are  given by the Green's functions $\mathcal{G}_{\zeta_+\zeta_+}$ and $\mathcal{G}_{\zeta_-\zeta_-}$, respectively.  
From the spin response function, the spin polarizability \cite{SOC_BosonsStringari2012b,2012EL.....9956008L} can be obtained, and at the mean-field level, we get
\begin{eqnarray}
\chi_{M}=\mathcal{G}_{0,\zeta_-\zeta_-}(q_x\to 0)/\rho=\frac{2}{\Omega-\Omega_{c,\mathrm{mf}}},
\end{eqnarray} 
which diverges at the mean-field phase transition point.

An important quantity characterizing  superfluidity is the superfluid density, which governs the total phase fluctuations. 
To get the superfluid density, we integrate out the $\varphi$,   $\zeta_-$ and $\zeta_+$ fields and obtain an effective theory of $\phi$ (see Appendix~\ref{app:meanfield}).
In the low energy and long wave length limit, we find 
\begin{eqnarray}\label{Eq:LEffective}
\mathcal{L}_{\mathrm{eff}}=(K\omega^2_n+\rho_{i} q^2_i)|\phi|^2,
\end{eqnarray}
where $K$ is the zero momentum  static density response function with its mean-field value being $1/g_+$, and $\rho_{y}=\rho_{z}=\rho$, $\rho_{x}=\rho [1-2k^2_0/(\Omega+2G_2) ]$ are the mean-field superfluid densities. 
From the effective Lagrangian, we see that the sound velocity is related to the  superfluid density through $c_i=\sqrt{K^{-1}\rho_i}$.

Note that the superfluid density in the $x$ direction vanishes when $\Omega=2k^2_0-2G_2$. 
Formally, for smaller $\Omega$, the superfluid density becomes negative, which means that a state with nonzero phase gradient, i.e. the plane wave phase, is energetically more favorable. 
In other words, a vanishing superfluid density indicates a second order phase transition from the zero momentum phase to the plane wave phase.

In \cite{PhysRevA.94.033635}, the superfluid density $\rho_x$ is calculated from the current-current correlation function, which can be written in terms of the transverse spin polarization $\langle \sigma_x\rangle$ and the excitation gap $\Delta$ as $\rho_{x}=\rho \left(1+2k^2_0\Omega\langle \sigma_x\rangle/\Delta^2\right)$.
Substituting the mean-field values $\Delta=\Delta_0$ and $\langle\sigma_x\rangle=-1$, we obtain from this the same result as given by the effective theory method above.  
Note that if the gap becomes larger or $\sigma_x$ is not fully polarized, the superfluid density $\rho_x$ will increase. 
  
\section{Beyond mean-field corrections}

To study the lowest order (one-loop) beyond mean-field corrections, we expand the Lagrangian density up to the fourth order of the fields,
\begin{widetext}
\begin{eqnarray}
\mathcal{L}_{\mathrm{fluct}}&=&
\frac{\zeta_+}{2}[(\nabla \phi)^2+(\nabla \varphi)^2]+\zeta_-\nabla\phi\nabla\varphi
  -\frac{\Omega\rho}{3}\varphi^4+\Omega\zeta_+\varphi^2-\frac{\Omega}{2\rho}\zeta^2_-\varphi^2-\frac{\Omega}{2} \bigg(\frac{\zeta_+ \zeta^2_-}{2\rho^2}-\frac{\zeta^2_-\zeta^2_+}{2\rho^3}-\frac{\zeta^4_-}{8\rho^3}\bigg)
  ,\nonumber\\
  &&
  -\frac{\zeta_+[(\nabla\zeta_+)^2 +(\nabla\zeta_-)^2]}{8\rho^2}
    -\frac{\zeta_-\nabla\zeta_+\nabla\zeta_-}{4\rho^2}
    +\frac{(\zeta^2_+ +\zeta^2_-)[(\nabla\zeta_+)^2+(\nabla\zeta_-)^2]+\nabla\zeta^2_+\nabla\zeta^2_-}{8\rho^3}.
\end{eqnarray}
 \begin{figure} 
	\includegraphics[width=\columnwidth]{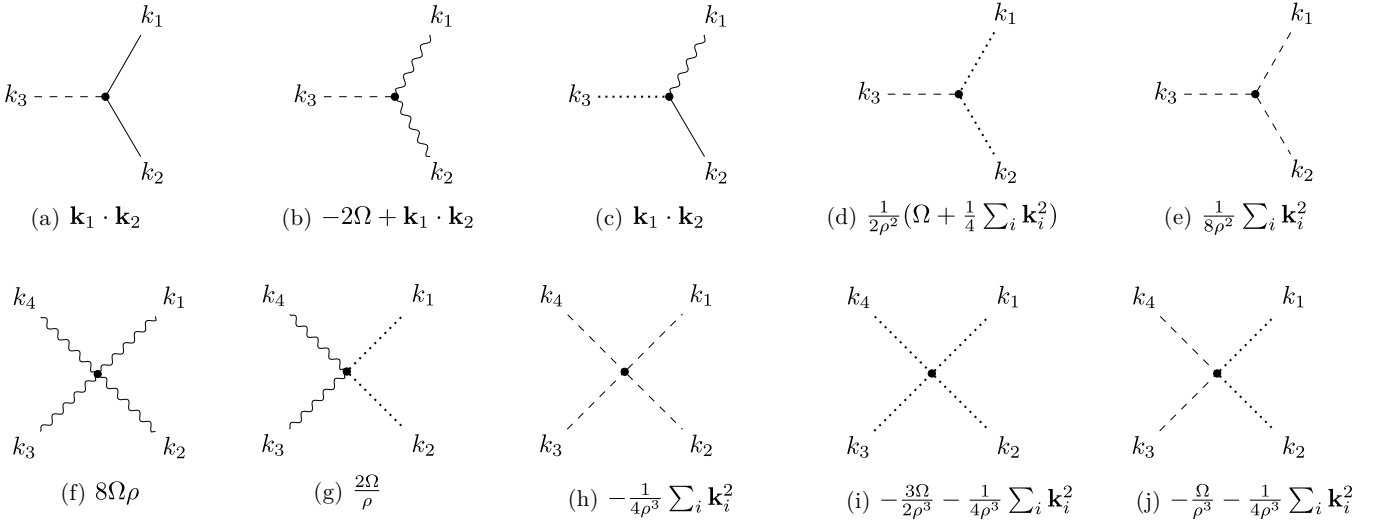} 
	\caption{Feynaman diagrams for interaction vertices.	
	}\label{Fig:Vertices}
\end{figure}  
\end{widetext}
The Feynman diagrams  corresponding to the vertices are given in Fig.~\ref{Fig:Vertices}.
Without the spin-orbit coupling, the one-loop corrections can be calculated analytically, and the results are given in Appendix~\ref{app:correctionswithoutsoc}. 
In the main text we focus on the more interesting situation with nonzero  spin-orbit coupling and calculate the one-loop corrections numerically. 
Since the parameter $G_2$ is small, we take it to be zero unless otherwise mentioned.

\subsection{Quantum depletion}

\begin{figure}
	\includegraphics[width=\columnwidth]{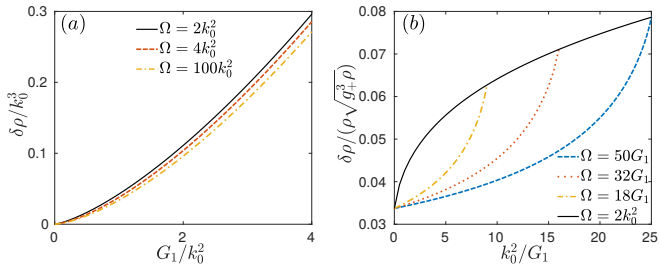}
	\caption{The quantum depletion as a function of the interaction (a) and spin-orbit coupling  (b) for different Raman fields $\Omega$. The solid lines correspond to the mean-field phase transition between the zero momentum and plane wave phases.
	}\label{Fig:quantum_depletion}
\end{figure}

Due to the quantum fluctuations, the condensate is depleted by a fraction of the total density. 
Up to the lowest order, the quantum depletion is given by (see Appendix~\ref{app:correctionswithoutsoc})
\begin{eqnarray}
\delta\rho=\rho \left(\langle \phi^2 \rangle +\langle \varphi^2 \rangle\right)+\frac{\langle \zeta^2_+\rangle +\langle \zeta^2_-\rangle}{4\rho}. 
\end{eqnarray}    
Fig.~\ref{Fig:quantum_depletion} shows the quantum depletion as a function of the interaction strength and spin-orbit coupling for different transverse fields. 
The quantum depletion increases with the interaction strength. 
We find that it also increases with the spin-orbit coupling strength, which is consistent with previous results \cite{PhysRevLett.109.025301,CuiZhouPRA2013}.
As Fig.~\ref{Fig:quantum_depletion} shows, the quantum depletion increases with decreasing $\Omega$, which means that the quantum fluctuations are enhanced as the system approaches the phase transition point.
 
\subsection{Lee-Huang-Yang correction and chemical potential shift}

\begin{figure}
	\includegraphics[width=\columnwidth]{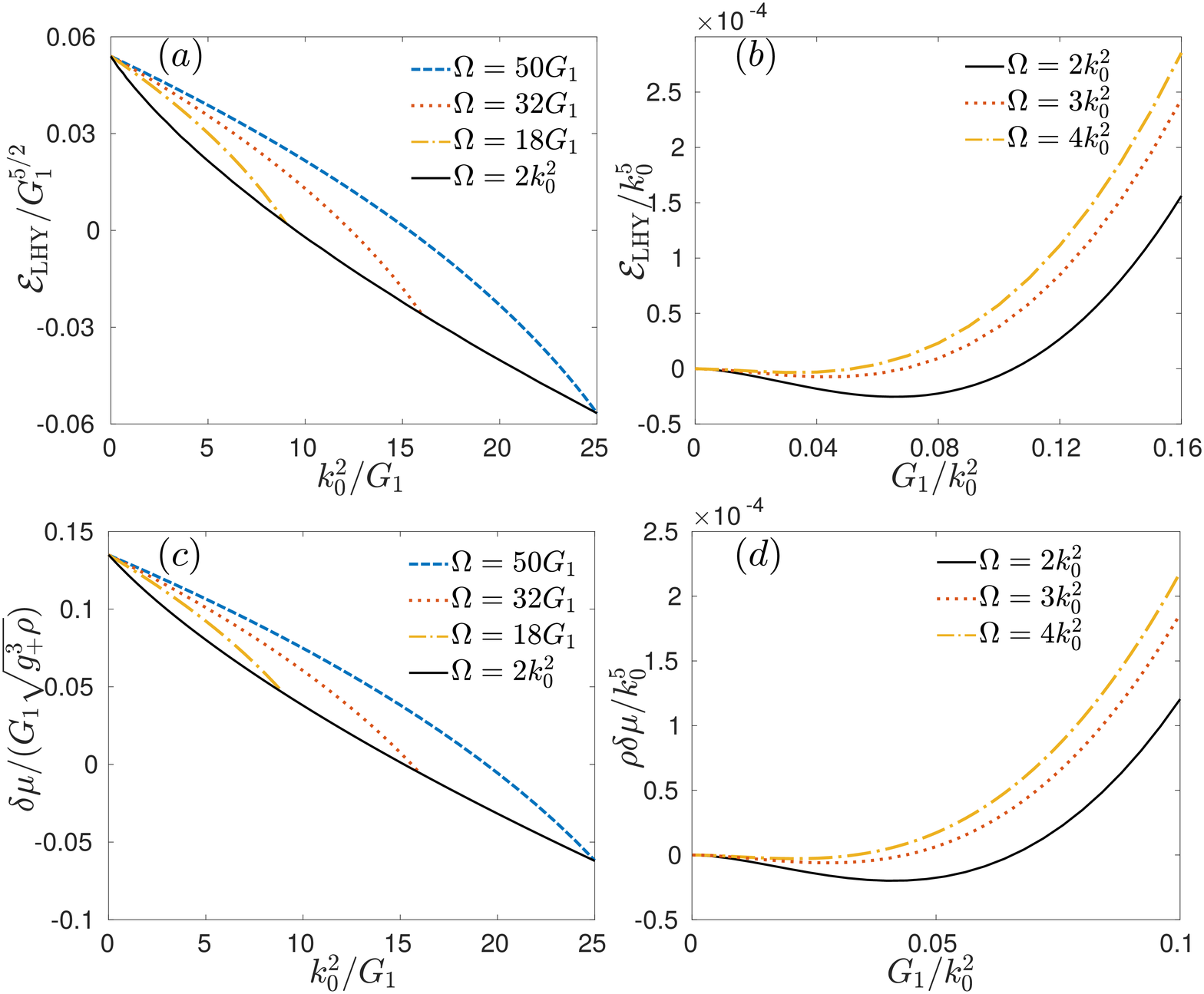}
	\caption{The LHY correction [(a) and (b)] and the chemical potential shift [(c) and (d)] as a function of the spin-orbit coupling  and interaction  for different Raman fields $\Omega$. The value $\Omega=2k^2_0$ corresponds to the mean-field phase transition.
	}\label{Fig:LHY}
\end{figure}

We study the correction to the mean-field energy density, which is known as the Lee-Huang-Yang (LHY) correction \cite{LHY}  $\mathcal{E}_{\mathrm{LHY}}$, and can be viewed as the zero point energy of the excitations \cite{RevModPhys.76.599}.
With increasing $k_0$, the phonon mode softens, and therefore the zero point energy decreases. 
Fig.~\ref{Fig:LHY} (a) shows this behavior clearly. 
Remarkably, we find that $\mathcal{E}_{\mathrm{LHY}}$ becomes negative for large enough spin-orbit coupling. 
This leads to a non-monotonic dependence of $\mathcal{E}_{\mathrm{LHY}}$ on $G_1$: If we fix $k_0$ and increase $G_1$ from zero, then for small $G_1$ (large $k^2_0/G_1$), the LHY correction decreases from zero to negative; increasing $G_1$ further, the LHY correction will increase since it becomes positive for small $k^2_0/G_1$. 
The non-monotonic behavior of $\mathcal{E}_{\mathrm{LHY}}$ is most clearly seen at the phase transition point, see Fig.~\ref{Fig:LHY} (b).

\begin{figure}
	\includegraphics[width=\columnwidth]{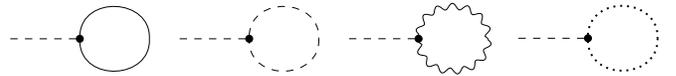}
	\caption{Feynman diagrams that determine the chemical potential shift $\delta \mu$. These tadpole diagrams are canceled by the chemical potential shift and therefore do not contribute to the one-loop self-energy. For notation see Fig.~\ref{Fig:G0}.  }\label{Fig:chemical_potential_shift}
\end{figure}

We then  calculate the correction to the chemical potential, which is given by the tadpole diagrams shown in Fig.~\ref{Fig:chemical_potential_shift}. 
The numerical results of $\delta\mu$ are shown in Figs.~\ref{Fig:LHY} (c) and (d). 
As the LHY correction, the chemical potential shift decreases with increasing of $k_0$ and depends non-monotonically on $G_1$. 
This is expected, because the chemical potential shift can also be obtained as the first order derivative of the LHY energy with respect to the density.

\subsection{ Superfluid density, phase boundary shift, and spin polarizability}  

\begin{figure}
	\includegraphics[width=\columnwidth]{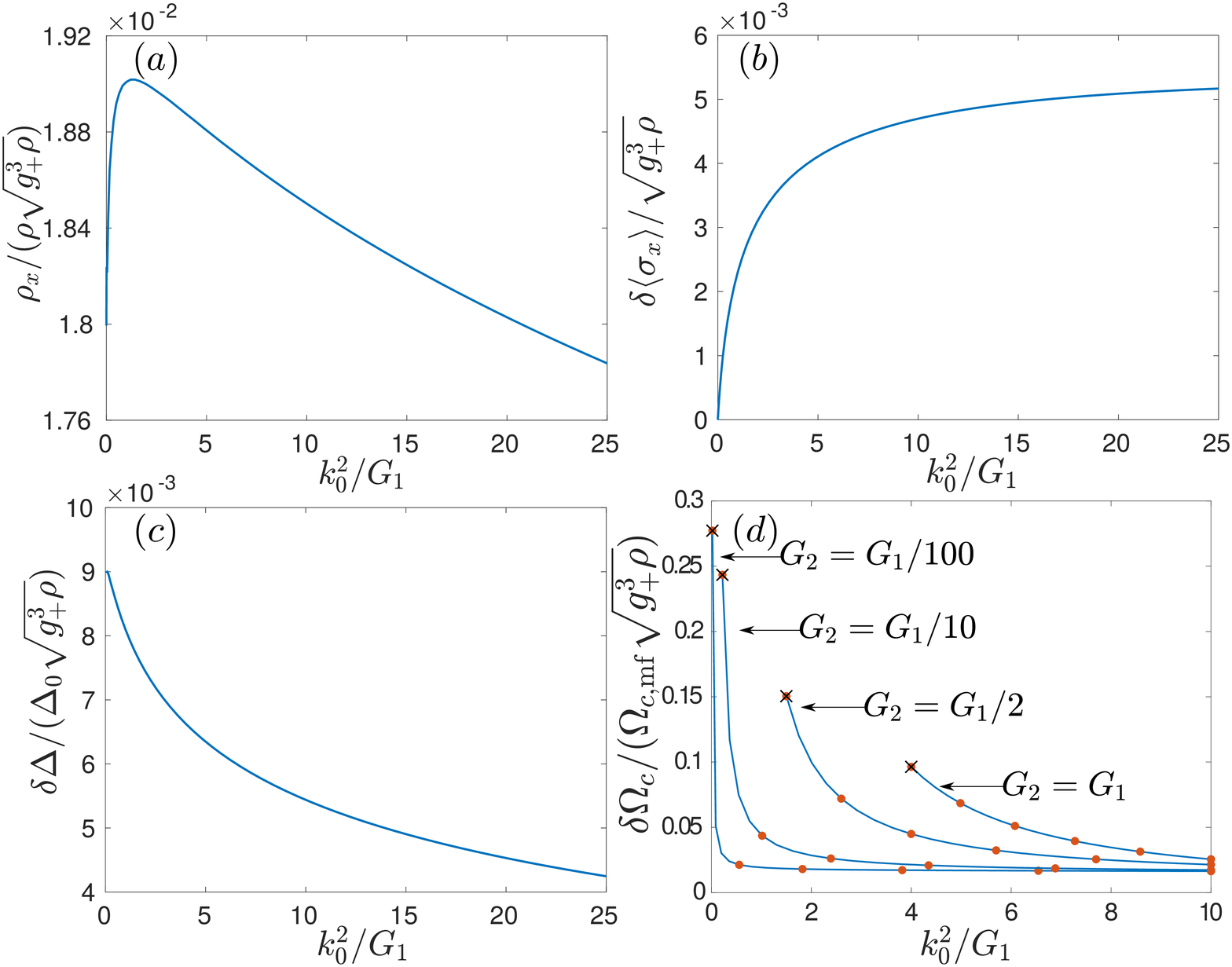}
	\caption{The superfluid density (a), the deviation of the transverse spin polarization (b), and the spin excitation gap (c) at the phase transition point. The phase boundary shift as a function of $k^2_0/G_2$ for different $G_2/G_1$ (d). Here $\delta \Omega_c=\Omega_{c,\mathrm{mf}}-\Omega_c$, with $\Omega_c$ being the corrected phase boundary. The solid lines show the results determined by the one-loop superfluid density, and the dots present the results by minimizing the ground state energy $\mathcal{E}_{\mathrm{mf}}+\mathcal{E}_{\mathrm{LHY}}$. The crosses denote the mean-field critical $k_{0,c}$ below which the plane wave phase is preempted by the stripe phase \cite{SOC_BosonsStringari2012a}.}
	\label{Fig:Dxx}
\end{figure}

To get the correction to the superfluid density, we first calculate the one-loop self-energy and then integrate out the massive fields $\varphi$, $\zeta_-$ and $\zeta_+$ to get the effective Lagrangian of the total phase fluctuations. 
The superfluid density in the $x$ direction is found to be
\begin{eqnarray}\label{Eq:supefluidensity_oneloop}
\rho_{x}&=&\rho\left[1-\frac{2k^2_0}{\Omega+2G_2-2\rho\Sigma_{\zeta_-\zeta_-}(0)}\right], 
\end{eqnarray}
where $\Sigma_{\zeta_-\zeta_-}(0)$ is the self-energy at zero frequency and momentum. There is no correction to $\rho_y$ and $\rho_z$ at zero temperature, consistent with the general result of superfluid density in Galilean invariant superfluids \cite{Leggett}.
 
Our numerical calculations show that $\Sigma_{\zeta_-\zeta_-}(0)$ is nonzero at the mean-field transition point. Consequently, the superfluid density also becomes nonzero at $\Omega=\Omega_{c,\mathrm{mf}}$, see Fig.~\ref{Fig:Dxx} (a).
Physically, this can be explained by the decrease of the transverse polarization $\langle\sigma_x \rangle$ and the increase of the spin gap $\Delta$. 
Because of the spin-orbit coupling, the spin of excited particles is not perfectly along the $x$ direction, and therefore the magnitude of the transverse spin polarization is reduced. 
Up to the lowest order, the deviation of spin polarization is (see Appendix~\ref{app:correctionswithoutsoc})  
\begin{eqnarray}
\delta\langle \sigma_x\rangle= 2\langle \varphi^2 \rangle+\frac{\langle \zeta^2_-\rangle}{2\rho^2}.
\end{eqnarray}
We plot the numerical result of $\delta\langle\sigma_x\rangle $ in Fig.~\ref{Fig:Dxx} (b). 
Another quantity that determines $\rho_x$ is the excitation gap. We obtain from the one-loop self-energy the correction to the mean-field gap and find it is positive, see Fig.~\ref{Fig:Dxx} (c). 
Combining the behavior of $\delta \langle\sigma_x\rangle$ and $\Delta$, the non-monotonic dependence of $\rho_{x}$ on $k^2_0$ can be explained: The superfluid density increases with increasing $\delta\langle\sigma_x\rangle$ and $\Delta$, and with increasing $k_0$, $\delta\langle\sigma_x\rangle$ increases but $\Delta$ decreases. 
As a result, the superfluid density first increases and then decreases with increasing the spin-orbit coupling strength. 
 
As we have explained before (see also Appendix~\ref{app:phaseboundary}), the phase transition between the zero momentum and plane wave phases is characterized by the vanishing superfluid density, so Eq.~\eqref{Eq:supefluidensity_oneloop} means that the phase transition point is shifted by quantum fluctuations. 
The new phase boundary is determined through
\begin{eqnarray}
\Omega_c+2G_2-2\rho\Sigma_{\zeta_-\zeta_-}(0)=2k^2_0,
\end{eqnarray} 
where $\Sigma_{\zeta_-\zeta_-}(0)$ should be evaluated at $\Omega_c$. 
The solid lines in Fig.~\ref{Fig:Dxx} (d) show the relative phase transition shift as a function of $k^2_0/G_1$ for different $G_2/G_1$.
The shift becomes larger with decreasing $k^2_0/G_1$ and 
reaches its maximum at a critical spin-orbit coupling strength $k_{0,c}$, below which the plane wave phase is preempted by the stripe phase \cite{SOC_BosonsStringari2012a}. We plot the phase boundary shift for $k_0$ larger than the mean-field critical value  $k_{0,c}=\sqrt{2G_2 (1+G_2/G_1)}$  \cite{SOC_BosonsStringari2012a}. It is possible that the mean-field critical spin-orbit coupling strength is shifted by  quantum fluctuations, but this is beyond the scope of this paper and we expect that it does not change the results presented in Fig.~\ref{Fig:Dxx} (d)  qualitatively.  
We also calculate the phase boundary by minimizing the ground state energy $\mathcal{E}_{\mathrm{mf}}+\mathcal{E}_{\mathrm{LHY}}$. 
The technical details are given in Appendix~\ref{app:phaseboundary}, and the phase boundary shifts obtained in this way are presented by the dots in Fig.~\ref{Fig:Dxx} (d). As can be seen, the two methods predict the same results.

 \begin{figure}
	\includegraphics[width=\columnwidth]{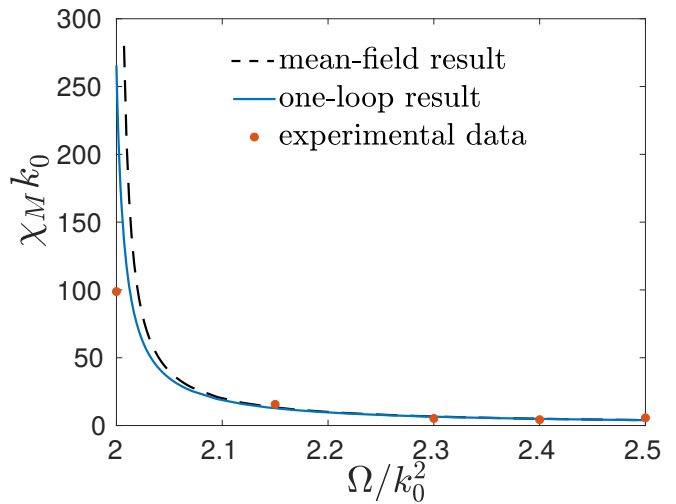}
	\caption{The spin polarizability as a function of $\Omega$. The experimental data are taken from \cite{PhysRevLett.109.115301}. To simulate the experiment, we use $k^2_0=4.2G_1$, $\sqrt{g^3_+\rho}=0.2$, and $G_2/G_1= 10^{-3}$ in the one-loop calculation.}
	\label{Fig:chi_M}
\end{figure}
The self-energy $\Sigma_{\zeta_-\zeta_-}(0)$ also gives a correction to the spin polarizability,
\begin{eqnarray}
\chi_{M}=\frac{2}{\Omega-\Omega_{c,\mathrm{mf}}-2\rho \Sigma_{\zeta_-\zeta_-}(0)}, 
\end{eqnarray}
which diverges at the corrected phase boundary but becomes finite at the mean-field phase transition point. We have checked numerically that around $\Omega_c$, the dependence of the self-energy $\Sigma_{\zeta_-\zeta_-}(0)$ on $\Omega$ is weak, and therefore $\chi_M$ diverges as  $1/(\Omega-\Omega_c)$ close to the phase boundary, as predicted by the mean-field theory.
The spin polarizability has been measured \cite{PhysRevLett.109.115301}, and it seems that our one-loop result agrees better with the experimental data than the mean-field theory, see Fig.~\ref{Fig:chi_M}. 
However, the current experimental data cannot lead to a decisive conclusion and future experiments are required to verify our prediction.

\subsection{Sound velocity and damping rate}
 
Using the one-loop results for the static density response $K^{-1}$ and the superfluid density $\rho_{x}$, we obtain the quantum corrected sound velocity in the $x$ direction, $c_x=\sqrt{K^{-1}\rho_x}$. At the corrected phase transition point, the sound velocity $c_x$ vanishes because of the vanishing superfluid density $\rho_x$. This is different from the result in \cite{PhysRevA.96.013625}, where a nonzero sound velocity at the phase boundary has been predicted within the Hartree-Fock-Bogoliubov-Popov approximation.

\begin{figure}
	\includegraphics[width=\columnwidth]{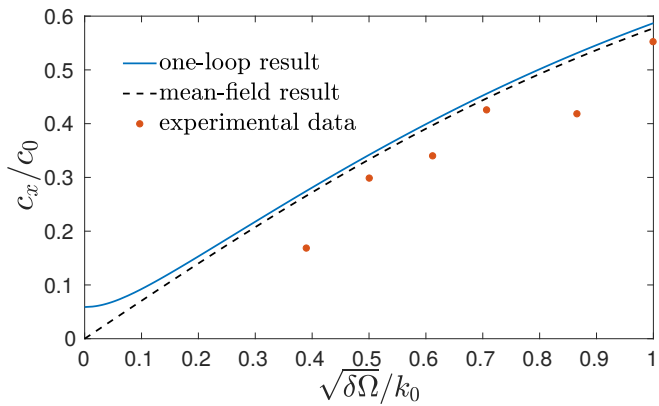}
	\caption{The sound velocity $c_x$ along the $x$ direction ($c_0=\sqrt{G_1}$). Here $\delta\Omega=\Omega-\Omega_{c,\mathrm{mf}}$ is the deviation of $\Omega$ from the mean-field phase transition value. The experimental data are taken from \cite{PhysRevLett.114.105301}, and the parameters we use to calculate the one-loop result  are $k^2_0=5.2G_1$, $\sqrt{g^3_+\rho}=0.18$, and $G_2/G_1=10^{-3}$, which correspond to the experiment in \cite{PhysRevLett.114.105301}. }
	\label{Fig:c_x}
\end{figure}

Since the sound velocity goes to zero slower than the superfluid density, it is easier to detect the beyond mean-field effects through the measurement of the sound velocity.  In Fig.~\ref{Fig:c_x} we plot the $c_x$ against $\sqrt{\delta \Omega}/k_0$, with $\delta\Omega=\Omega-\Omega_{c,\mathrm{mf}}$. 
For typical experimental parameters \cite{2011Natur.471...83L,PhysRevLett.109.115301,PhysRevLett.114.105301}, 
the one-loop prediction deviates clearly from the mean-field behavior when $\sqrt{\delta\Omega}/k_0<0.1$.
The sound velocity has been measured \cite{PhysRevLett.114.105301}, but the parameters are not close enough to the phase transition point. 
However, our prediction should be observable with current experimental methods.

Finally, we calculate the damping rate of phonons, for details see Appendix~\ref{app:damping}. 
At zero temperature, the damping is due to the Beliaev process \cite{Beliaev}, i.e., an excitation decays into two with lower energy. In the small momentum limit ($q_y,q_z\ll \sqrt{G_1}$ and $q_x\ll\sqrt{\Omega-2k^2_0}$),
we find
\begin{eqnarray}
\gamma_\mathrm{B}
&=&\frac{3q^5}{640\pi\rho}\left[1-\frac{2\Omega k^ 2_0\cos^2{\theta}}{(\Omega+2G_2)^2}\right]^2
\sqrt{1+\frac{2k^2_0\sin^2{\theta}}{\Omega-\Omega_{c,\mathrm{mf}}}}, \nonumber\\
\end{eqnarray}
which coincides with the result obtained in \cite{Beliaev_SOC}. The Beliaev damping is strongly suppressed along the direction of the spin-orbit coupling.

At finite temperature, the Landau damping \cite{HOHENBERG1965291} arises because the phonon couples to thermal excitations. 
The Landau damping is experimentally more relevant since it is responsible for damping in trapped Bose gases \cite{1997PhLA..235..398P,PhysRevLett.79.4056,PhysRevLett.80.2269}. 
In the low temperature and small momentum limit ($c_{\theta}q\ll T\ll \Omega-2k^2_0$), we obtain 
\begin{eqnarray}\label{Eq:LandauDamping}
\gamma_\mathrm{L}=\frac{3\pi^3 q T^4}{40\rho c^4_{\theta}}\left[1-\frac{2\Omega k^ 2_0\cos^2{\theta}}{(\Omega+2G_2)^2}\right]^2\sqrt{1+\frac{2k^2_0\sin^2{\theta}}{\Omega-\Omega_{c,\mathrm{mf}}}}. \nonumber\\
\end{eqnarray}
Because of the extra $c_\theta$ dependence, 
the Landau damping rate, unlike the  Beliaev decay, is not suppressed in the direction of spin-orbit coupling, which means that the Landau process is the dominant damping mechanism even for uniform systems at very low temperature.

\section{Conclusions} 

We calculate systematically the one-loop corrections to a spin-orbit coupled Bose-Einstein condensate. 
We find that quantum fluctuations cause quantitative modifications to the superfluid density, spin polarizability, sound velocity, and damping rate.   
The quantum depletion increases while the LHY energy decreases with the transverse field in the zero momentum phase.
The phase boundary between the plane wave and zero momentum phases is shifted to a smaller transverse field.
The superfluid density vanishes and the spin polarizability diverges at the one-loop phase transition point. 
But at the mean-field phase boundary, the spin polarizability remains finite, consistent with an experimental measurement \cite{PhysRevLett.109.115301}. 
We also point out that the beyond mean-field corrections may be detected through the measurement of the sound velocity, and give the parameter regime in which the deviation from the mean-field behavior is visible.  
We calculate the Beliaev and Landau damping rates and identify the Landau damping as the dominant mechanism of quasiparticle decay. 
Our results show that the spin-orbit coupling leads to, even for moderate interactions, quantum fluctuations strong enough to make detectable modifications to the properties of a macroscopic quantum state such as a Bose-Einstein condensate. 
The results can be readily tested in ultracold quantum gases, and in the future, in spin-orbit coupled Bose-Einstein condensates realized in other systems.

\section{Acknowledgements}

This work was supported by theAcademy of Finland under Projects No.  303351, No. 307419, No. 318987, and by the European Research
Council (ERC-2013-AdG-340748-CODE). L.L. would like to acknowledge the Aalto Centre for Quantum Engineering for support.

\appendix

\section{Mean-field results}\label{app:meanfield}
 
In this section we present the mean-field results of the excitation energy, density and spin response function, and superfluid density  with some detailed derivations.

\begin{widetext}

\subsection{Excitation energy}
The excitation energy is determined by $\det \mathcal{G}^{-1}_{0}=0$, which gives
\begin{eqnarray}
&&\varepsilon^2_{\mathrm{ph}}(\mathbf{q})=\frac{a(\mathbf{q}) - \sqrt{ a^2(\mathbf{q})-b(\mathbf{q}) }}{2},\\
&&\varepsilon^2_{\mathrm{sp}}(\mathbf{q})=\frac{a(\mathbf{q}) + \sqrt{ a^2(\mathbf{q})-b(\mathbf{q}) }}{2},
\end{eqnarray}
where $a(\mathbf{q})=\mathcal{A}(\mathbf{q}) \mathcal{B}(\mathbf{q}) + \mathcal{C}(\mathbf{q}) \mathcal{D}(\mathbf{q}) + 2k^2_0q^2_x$, $b(\mathbf{q})=4[\mathcal{A}(\mathbf{q}) \mathcal{D}(\mathbf{q})-k^2_0q^2_x][\mathcal{B}(\mathbf{q}) \mathcal{C}(\mathbf{q}) -k^2_0q^2_x]$, $\varepsilon_{\mathrm{ph}}$ is the gapless phonon mode, and $\varepsilon_{\mathrm{sp}}$ is the gapped mode which is dominated by spin excitations. 
In the small momentum limit,
\begin{eqnarray}
&&\varepsilon_{\mathrm{sp}}=\Delta_0+
\frac{q^2}{2m_{\mathrm{sp}}},\\
&&\varepsilon_{\mathrm{ph}}=c_{\theta}q+d_{\theta}q^3,
\end{eqnarray}

where
\begin{eqnarray}
&&\Delta_0=\sqrt{\Omega(\Omega+2G_2)},\\
&& m^{-1}_{\mathrm{sp}}=\frac{(G_2 + \Omega) (2 G_2 + \Omega) + 
	2 k^2_0 (\Omega+G_1+2 G_2) \cos^2{\theta}}{\sqrt{\Omega} (2 G_2 + \Omega)^{3/2}},\\
 &&c_{\theta}=c_0\sqrt{1-\frac{2k^2_0\cos^2{\theta}}{\Omega+2G_2}},\\
 &&d_{\theta}=\frac{1}{8c_{\theta}}\left(
 1-\frac{4k^2_0[(\Omega+2G_2)(\Omega^2+(G_1+3G_2)\Omega+2(G_1+G_2)^2)\cos^2{\theta}-k^2_0(\Omega+2(G_1+G_2))^2\cos^4{\theta}]}{\Omega(\Omega+2G_2)^3} 
 \right),
\end{eqnarray}
with $c_0=\sqrt{G_1}$ being the usual Bogoliubov sound velocity for a weakly interacting single component Bose-Einstein condensate. In the absence of spin-orbit coupling, the sound velocity is the same as $c_0$. 
In the presence of spin-orbit coupling, it depends on $\theta$, which is the angle between the momentum $\mathbf{q}$ and direction of the spin-orbit coupling. 
When $\Omega=2k^2_0-2G_2$, the sound velocity along the $x$ direction becomes zero, and the phonon dispersion along the $x$ direction becomes quadratic,
\begin{eqnarray}
\varepsilon_{\mathrm{ph}}=\sqrt{G_1 q^2_{\parallel}+\frac{1}{4}q^4_{\parallel}+\frac{G_2 q^4_x}{4 G_2 + 2 \Omega} -\frac{(G_1 + G_2)[2 (G_1 + 
	G_2) + \Omega] }{2 
	\Omega (2 G_2 + \Omega)}q^2_{\parallel}q^2_{x} },
\end{eqnarray}
with $q^2_{\parallel}=q^2_y+q^2_z$.

Knowing the low energy dispersion relation of the phonons, 
we can define the momentum region in which the dispersion is linear. 
When the momentum is along the $x$ direction, by requiring $c_x q_x \gg d_x q^3_x$, we find the condition
\begin{eqnarray}\label{Eq:Linear1}
q_x\ll \sqrt{\Omega-2k^2_0}.
\end{eqnarray}
When the momentum is along the $y$ or $z$ direction, the condition is 
\begin{eqnarray}\label{Eq:Linear2}
q_y, q_z\ll \sqrt{G_1}.
\end{eqnarray}
At finite temperature, the linear dispersion region also requires that the dispersion of the thermal excitations is linear, and this leads to the condition
\begin{eqnarray}\label{Eq:Linear3}
T\ll \Omega-2k^2_0.
\end{eqnarray}
These conditions are used in deriving the analytical expressions for Beliaev and Landau damping rates.

\subsection{Density and spin response functions}

In the modulus-phase representation,  
the density and spin response functions are  given by the Green's functions $\mathcal{G}_{\zeta_+\zeta_+}$ and $\mathcal{G}_{\zeta_-\zeta_-}$, respectively. So the spin polarizability defined in \cite{SOC_BosonsStringari2012b,2012EL.....9956008L} is simply given by $\mathcal{G}_{\zeta_-\zeta_-}(q_x\to 0)/\rho$, and at the mean-field level,
\begin{eqnarray}
\chi_{M}=\mathcal{G}_{0,\zeta_-\zeta_-}(q_x\to 0)/\rho=\frac{2}{\Omega+2 G_2-2k^2_0}.
\end{eqnarray}

The mean-field density and spin static structure factors are given by
\begin{eqnarray}
 S_d(\mathbf{q})=\int \mathrm{d}\omega \mathcal{G}_{0,\zeta_+\zeta_+}(\omega,\mathbf{q})/\rho,~~ S_s(\mathbf{q})=\int \mathrm{d}\omega \mathcal{G}_{0,\zeta_-\zeta_-}(\omega,\mathbf{q})/\rho.
\end{eqnarray}
We show the mean-field static structure factors for different spin-orbit coupling strength in Fig.~\ref{Fig:static_structure_factor}. 
As comparison, the contribution of the phonon branch are also shown.
Without spin-orbit coupling, the density and spin excitations are decoupled and the phonon branch does not contribute to the spin structure factor. 
In the presence of spin-orbit coupling, a density perturbation along the $x$ direction also induces a spin response and vice versa, so the density and spin structure factors are carried by both the phonon and gapped excitations. 
In the large momentum limit, the total static structure factors approach to 1 and the phonon branch contributes to one half.  
Remarkably, we find a peak in the total spin static structure factor. 
When the parameter approaches to the phase transition point, the peak becomes higher and its location moves to the zero momentum. 
By contrast, the  peak is not observed in the total density structure factor, although there is peak in the contribution of the phonon branch.
 
\begin{figure}
	\includegraphics[width=\columnwidth]{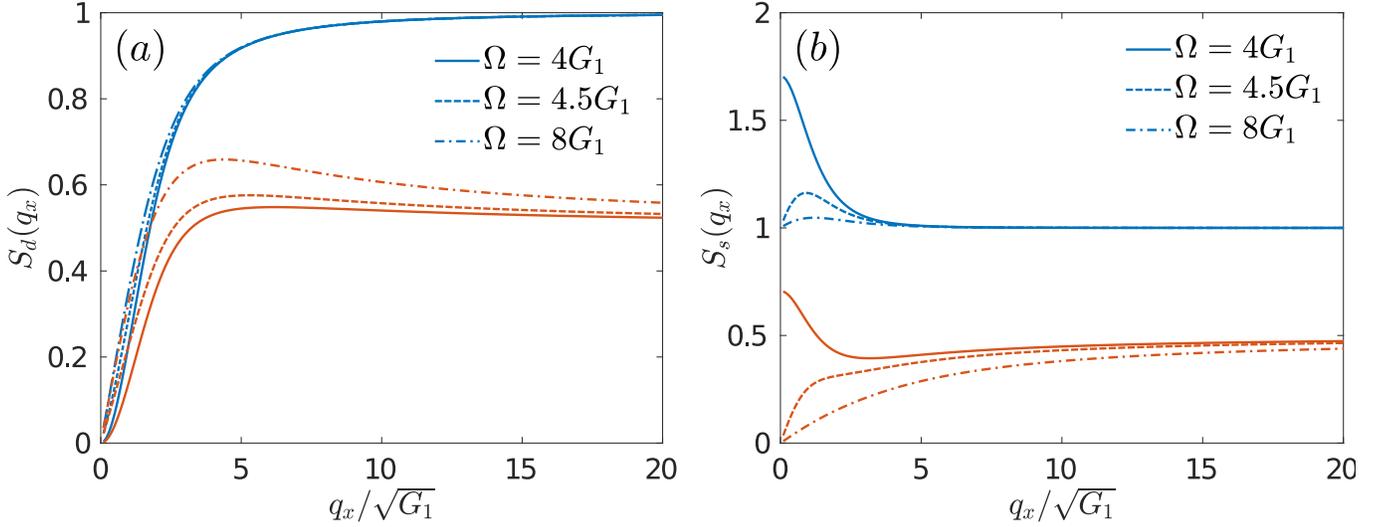}
 	\caption{The density (a) and spin (b) static structure factors (blue upper lines). Both approach unity in the large momentum limit. The contribution of the phonon branch are also shown (red lower lines). The interaction is taken to be $SU(2)$ invariant with $G_2=0$ and $G_1$ is taken to be unit. There is a peak in the spin structure factor.
 	}\label{Fig:static_structure_factor}
\end{figure}

\subsection{Superfluid density}
 
To get the superfluid density, we integrate out the $\varphi$ and $\zeta_-$ fields and obtain an effective theory of $\phi$ and $\zeta_+$
\begin{eqnarray}\label{Eq:EffetiveLagrangian}
\mathcal{L}_{\mathrm{eff}}&=&\frac{1}{2}[\phi,\zeta_+]\mathcal{G}^{-1}_{0,\mathrm{eff}}[\phi,\zeta_+]^{T},
\end{eqnarray}
where 
\begin{eqnarray}
\mathcal{G}^{-1}_{0,\mathrm{eff}}&=&\left[\begin{array}{cc}
\rho \mathbf{q}^2  & -\omega_n  \\ 
\omega_n & \frac{\mathbf{q}^2}{4\rho}+g_+ 
\end{array} 
\right]-
\left[\begin{array}{cc}
0 & i k_0 q_x \\ 
-i k_0 q_x & 0
\end{array} 
\right]
\left[\begin{array}{cccc} 
\rho \mathbf{q}^2+2\Omega\rho & -\omega_n  \\ 
\omega_n & \frac{\mathbf{q}^2}{4\rho}+g_- +\frac{\Omega}{2\rho}
\end{array} 
\right]^{-1}   
\left[\begin{array}{cc}
0 & i k_0 q_x \\ 
-i k_0 q_x & 0
\end{array} 
\right],               
\end{eqnarray} 
which in the low energy limit is
\begin{eqnarray}
\mathcal{G}^{-1}_{0,\mathrm{eff}}&=&\left[\begin{array}{cc}
\rho (\mathbf{q}^2 -\frac{2k^2_0}{\Omega +2G_2} q^2_x )  & -\omega_n  \\ 
\omega_n & \frac{\mathbf{q}^2}{4\rho}+g_+ 
\end{array} 
\right].
\end{eqnarray}
Integrating out the $\zeta_+$ field, we arrive at an effective Lagrangian of the phase fluctuation, and in the low energy and long wave length limit,
\begin{eqnarray}
\mathcal{L}_{\mathrm{eff}}= (K^{-1}\omega_n^2+\rho_i q^2_i)|\phi|^2,
\end{eqnarray}
where  $K$ is the zero momentum static density response function whose mean-field value is $1/g_+$, and the mean-field superfluid densities are
\begin{eqnarray}
\rho_x=\rho\left(1-\frac{2k^2_0}{\Omega+2G_2}\right),~\rho_y=\rho_z=\rho.
\end{eqnarray}
  
\section{Analytical results of one-loop corrections in the absence of spin-orbit coupling}\label{app:correctionswithoutsoc}

Without the spin-orbit coupling, we can calculate the one-loop corrections analytically. It is useful to calculate the following integral,
\begin{eqnarray}
&&I(\alpha,m^2;\beta,M^2;\gamma)=\int\frac{\mathrm{d}^dk}{(2\pi)^d}\frac{k^{\gamma}}{(k^2+m^2)^{\alpha}(k^2+M^2)^{\beta}}=\frac{2\pi^{d/2}}{(2\pi)^d\Gamma(d/2)}\int\mathrm{d}k \frac{k^{d-1+\gamma}}{(k^2+m^2)^{\alpha}(k^2+M^2)^{\beta}},\\
&&=\frac{2\pi^{d/2}}{(2\pi)^d\Gamma(d/2)}\int\mathrm{d}k \frac{\Gamma(\alpha+\beta)}{\Gamma(\alpha)\Gamma(\beta)}\int^1_0 \mathrm{d}x \frac{k^{d-1+\gamma}x^{\alpha-1} (1-x)^{\beta-1}}{[x(k^2+m^2)+(1-x)(k^2+M^2)]^{\alpha+\beta}},\\
&&=\frac{2\pi^{d/2}}{(2\pi)^d\Gamma(d/2)}
\frac{(M^2)^{\frac{d+\gamma}{2}-\alpha-\beta}}{\Gamma(\alpha+\beta)}\Gamma\left(\alpha+\beta-\frac{d+\gamma}{2}\right)\Gamma\left(\frac{d+\gamma}{2}\right){}_2F_1\left(\alpha,\alpha+\beta-\frac{d+\gamma}{2};\alpha+\beta;1-\frac{m^2}{M^2}\right),
\end{eqnarray}
where $_2F_1(a,b;c;z)$ is the hypergeometric function. 
To get the above result we have used dimensional regularization.

The condensate fraction is 
\begin{eqnarray}
|\langle\psi_\uparrow \rangle|^2+|\langle\psi_\downarrow \rangle|^2&=&\frac{1}{2}\left(|\langle \sqrt{\rho+\zeta_1}e^{i(\phi+\varphi)}\rangle|^2+|\langle -\sqrt{\rho+\zeta_2}e^{i(\phi-\varphi)}\rangle|^2\right)
\approx\rho -\frac{1}{4\rho}(\langle \zeta^2_+ \rangle+ \langle \zeta^2_- \rangle )- \rho(\langle \varphi^2 \rangle +\langle \phi^2 \rangle ),~
\end{eqnarray}
so the quantum depletion is
\begin{eqnarray}
\delta \rho &=&\frac{1}{4\rho}(\langle \zeta^2_+ \rangle+ \langle \zeta^2_- \rangle )+ \rho(\langle \varphi^2 \rangle +\langle \phi^2 \rangle )
=\frac{(g_+\rho)^{3/2}}{3\pi^2} + \frac{(g_-\rho)^{3/2}}{3\pi^2}\sqrt{1+x/2}
\left[(x+1)E\left(\frac{2}{x+2}\right)-x K\left(\frac{2}{x+2}\right) \right],~~~
\end{eqnarray}
with $x=\Omega/(g_-\rho)$ and $E(z)$ and $K(z)$ are the complete elliptic integral of the second and first kind, respectively. 
The quantum depletion increases with increasing $g_+$ and $g_-$, but decreases with increasing $\Omega$.

The transverse spin polarization is
\begin{eqnarray}
\langle \sigma_x\rangle=\frac{1}{\rho}\langle \psi^\dag \sigma_x \psi \rangle\approx -1+2\langle \varphi^2\rangle+\frac{1}{2\rho^2}\langle\zeta^2_- \rangle, 
\end{eqnarray}
so
\begin{eqnarray}
\delta \langle\sigma_x \rangle &=& 2\langle \varphi^2\rangle+\frac{1}{2\rho^2}\langle\zeta^2_- \rangle
= \frac{2(g_-\rho)^{3/2}}{3\rho \pi^2}\sqrt{1+x/2}
\left[(x+1)E\left(\frac{2}{x+2}\right)-x K\left(\frac{2}{x+2}\right) \right].
\end{eqnarray}

The Lee-Huang-Yang correction \cite{LHY} can be obtained as the zero point energy of the system \cite{RevModPhys.76.599}, and we find
\begin{eqnarray}\label{Eq:ELHY}
 \mathcal{E}_{\mathrm{LHY}}&=&\frac{8}{15\pi^2}g_+\rho^2\sqrt{g^3_+\rho}+\frac{(g_-\rho)^{5/2} \sqrt{1 + x/2} }{ 4\pi}
 {}_2F_1\left(-\frac{1}{2},\frac{3}{2};3;\frac{2}{2 + x}\right).
\end{eqnarray}
The first term in Eq.~\eqref{Eq:ELHY} is the same as the result for  a weakly-interacting spinless Bose gas \cite{LHY}. 
The second term comes from the spin excitation. 
The function ${} _2F_1\left(-\frac{1}{2},\frac{3}{2};3;\frac{2}{2 + x}\right)$ depends weakly on $x$, with ${} _2F_1\left(-\frac{1}{2},\frac{3}{2};3;1\right)=32/(15\pi)$ and ${} _2F_1\left(-\frac{1}{2},\frac{3}{2};3;0\right)=1$, so the second term increases with increasing $g_-$ and $\Omega$.
    
The chemical potential shift is given by the tadpole diagrams shown in Fig.~\ref{Fig:chemical_potential_shift}.   Evaluating the integrals, we find
  \begin{eqnarray}\label{Eq:deltamu}
  \delta\mu &=&\frac{4 g_+\rho\sqrt{g^3_+\rho}}{3\pi^2}+\frac{g_-\rho\sqrt{g^3_-\rho}\sqrt{1+x/2}}{3\pi^2}\left[(4+x)E\left(\frac{2}{2+x}\right)-x K\left(\frac{2}{2+x}\right)
  \right],
  \end{eqnarray}
which increases with $g_+$, $g_-$, and $\Omega$. 
Another way to calculate the chemical potential shift is to take derivative of the LHY energy density $\mathcal{E}_{\mathrm{LHY}}$ with respect to $\rho$, $\delta\mu =\partial_\rho \mathcal{E}_{\mathrm{LHY}}$, and the result is the same as Eq.~\eqref{Eq:deltamu}.

The correction to $K^{-1}$ is given by $\Sigma_{\zeta_+\zeta_+}(0)$, and in the absence of spin-orbit coupling, 
\begin{eqnarray}
 \delta K^{-1}=-\Sigma_{\zeta_+\zeta_+}(0) 
 &=&\frac{2 g_+ \sqrt{g^3_+\rho}}{\pi^2}+\frac{g_- \sqrt{g^3_-\rho}\sqrt{1+x/2}}{2\pi}{} _2F_1\left(-\frac{1}{2},-\frac{1}{2};1;\frac{2}{2+x}\right).
\end{eqnarray}
Note that $\delta K^{-1}$ can be related to $\delta\mu$ through $\delta K^{-1}=\partial_\rho \delta\mu$.
 
To calculate the correction to the mean-field excitation gap $\Delta_0$, we need to compute the self-energies $\Sigma_{\varphi\varphi}(\Delta_0)$, $\Sigma_{\zeta_- \zeta_-}(\Delta_0)$, and $\Sigma_{\varphi\zeta_-}(\Delta_0)$, which can also be done analytically in the absence of the spin-orbit coupling, and we find that the one-loop correction to the gap is zero. 
In the presence of spin-orbit coupling, we calculate the self-energies numerically, and find the one-loop correction increases the gap slightly, see the main text.

 \section{Phase boundary between the plane wave and zero momentum phases: the effect of the LHY energy}\label{app:phaseboundary}
 
In this section we study the phase boundary between the zero momentum and plane wave phases by minimizing the ground state energy. As we will show, this also provides another way to calculate the superfluid density.
 
We only consider the plane wave and zero momentum phases, and in general the field operator can be written as
\begin{eqnarray}\label{Eq:PW_field_parameterization}
 	\psi=e^{i \phi+i k_1 x}\left[\begin{array}{c}
 		\sqrt{\rho+\zeta_1}\cos{\alpha} e^{i\varphi}\\ 
 		- \sqrt{\rho+\zeta_2}\sin{\alpha} e^{-i\varphi}
 	\end{array} \right],
\end{eqnarray} 
where we have introduced the phase fluctuations $\phi$ and $\varphi$, and the density fluctuations $\zeta_1$ and $\zeta_2$, which are simply set to be zero in the mean-field approximation. 
The parameters $k_1$ and $\alpha$ should be determined  by minimizing the ground state energy, and $k_1\ne 0$ characterizes the plane wave phase while $k_1=0$ gives the zero momentum phase.

Substituting Eq.~\eqref{Eq:PW_field_parameterization} to the Lagrangian density
\begin{eqnarray}
\mathcal{L}&=&\psi^\dagger(\partial_\tau+h_0-\mu)\psi
+\frac{g_+}{2} (\psi^\dagger\psi)^2+\frac{g_-}{2}(\psi^\dagger\sigma_z\psi)^2,
\end{eqnarray}
and up to the  quadratic order of the fluctuations, we find
\begin{eqnarray}\label{Eq:Lagrangian_PW}
\mathcal{L}&=&
\frac{k^2_1+k^2_0}{2}\rho-k_0 k_1 \rho \cos{2\alpha}  -\frac{\Omega}{2} \rho\sin{2\alpha}+\frac{g_+ + g_- \cos^2{2\alpha}}{2}\rho^2 -\mu\rho \nonumber\\
&&+\left( \frac{k^2_1+k^2_0}{2} -\frac{\Omega}{2 \sin{2\alpha}} +G_1 -\mu \right) \zeta_+ +\left( \frac{\Omega \cos{2\alpha}}{2 \sin{2\alpha}} -k_0 k_1 +G_2 \cos{2\alpha}\right) \zeta_-\nonumber\\
&&+\frac{1}{2} [\phi,\zeta_+,\varphi,\zeta_-]\mathcal{G}^{-1}_{0}[\phi,\zeta_+,\varphi,\zeta_-]^{T},
\end{eqnarray}
where $\zeta_+=\zeta_1 \cos^2{\alpha}+\zeta_2 \sin^2{\alpha}$,  $\zeta_-=\zeta_1 \cos^2{\alpha}-\zeta_2 \sin^2{\alpha}$, and $\mathcal{G}^{-1}_{0}$ in the momentum and frequency representation reads
\begin{eqnarray}\label{Eq:Green_PW}
\mathcal{G}^{-1}_{0}(i\omega,\mathbf{k})=\left[\begin{array}{cccc}
\rho k^2  & -\omega -i k_1 k_x & \rho k^2 \cos{2\alpha} & i k_0 k_x \\ 
\omega+i k_1 k_x & \frac{k^2}{4\rho \sin^2{2\alpha}}+g_+ +\frac{\cos^2{2\alpha}}{2\sin^3{2\alpha}}\frac{\Omega}{\rho} & -i k_0 k_x & -\frac{\Omega \cos{2\alpha}}{2\rho\sin^3{2\alpha}} -\frac{\cos{2\alpha}}{4\rho \sin^2{2\alpha}}k^2\\ 
\rho k^2 \cos{2\alpha} & i k_0k_x & \rho k^2+2\Omega\rho \sin{2\alpha} & -\omega -i k_1 k_x  \\ 
-i k_0k_x   & -\frac{\Omega \cos{2\alpha}}{2\rho\sin^3{2\alpha}}-\frac{\cos{2\alpha}}{4\rho \sin^2{2\alpha}}k^2 &  \omega +i k_1 k_x & \frac{k^2}{4\rho \sin^2{2\alpha}}+g_- +\frac{1}{2\sin^3{2\alpha}}\frac{\Omega}{\rho} 
\end{array} 
\right]. 
\end{eqnarray}

We choose the renormalization condition $\langle \zeta_+ \rangle =\langle \zeta_- \rangle=0$, which gives two conditions at the mean-field level
\begin{eqnarray}
&&\frac{k^2_1+k^2_0}{2} -\frac{\Omega}{2 \sin{2\alpha}} +G_1 -\mu =0,\label{Eq:Cond1}\\
&&  \frac{\Omega \cos{2\alpha}}{2 \sin{2\alpha}} -k_0 k_1 +G_2\cos{2\alpha}=0.\label{Eq:Cond2}
\end{eqnarray}
The first condition Eq.~\eqref{Eq:Cond1} determines the mean-field chemical potential and the second condition Eq.~\eqref{Eq:Cond2} gives a relation between $\alpha$ and $k_1$. Note that for small $k_1$, we have $\cos{2\alpha} \propto k_1 $. 

The mean-field energy density is given by the first line in Eq.~\eqref{Eq:Lagrangian_PW},
 \begin{eqnarray}
\mathcal{E}_{\mathrm{mf}}&=&
\frac{k^2_1+k^2_0}{2}\rho-k_0 k_1 \rho \cos{2\alpha}  -\frac{\Omega}{2} \rho\sin{2\alpha} +\frac{g_+ + g_- \cos^2{2\alpha}}{2}\rho^2.
\end{eqnarray}
Note that Eq.~\eqref{Eq:Cond2} can also be obtained by minimizing the energy with respect to $\alpha$.
In \cite{SOC_BosonsStringari2012a}, a relation between $k_1$ and $\alpha$ is obtained by minimizing the energy with respect to $k_1$, which leads to $\alpha=\arccos{(k_1/k_0)}/2$. 
This relation and Eq.~\eqref{Eq:Cond2} determine the mean-field value of $\alpha$ and $k_1$ and therefore the mean-field phase boundary, which are the same as the results in \cite{SOC_BosonsStringari2012a}.
However, $\alpha=\arccos{(k_1/k_0)}/2$ no longer holds when the LHY energy is taken into account because in this case there will be extra  contribution to the energy density  depending on $k_1$. 
In contrast, Eq.~\eqref{Eq:Cond2} is still valid up to at least one-loop since there is no one-loop correction proportional to $\zeta_-$ and therefore  $\langle \zeta_-\rangle=0$ leads to the same condition. 
Therefore, to include the effects of the LHY energy, we should utilize the condition Eq.~\eqref{Eq:Cond2} instead of the form used in \cite{SOC_BosonsStringari2012a}.

Using Eq.~\eqref{Eq:Cond2}, we can rewrite $\mathcal{E}_{\mathrm{mf}}$ in terms of $k_1$, and then we can view the resultant expression as a Landau functional in terms of the `order parameter' $k_1$. 
The disordered phase corresponds to the zero momentum phase while the ordered phase is the plane wave phase. 
Technically, it is simpler to use $x\equiv \cos{2\alpha}$ as the order parameter (because $x\propto k_1$ for small $k_1$) and  we have
\begin{eqnarray}\label{Eq:MFEnergy}
\mathcal{E}_{\mathrm{mf}}
&=&\frac{k^2_0}{2}\rho -\frac{\Omega}{2 \sqrt{1-x^2}}\rho +\frac{g_+\rho^2}{2}+\frac{\rho}{2k^2_0} \left[\frac{\Omega x}{2 \sqrt{1-x^2}} +g_-\rho x\right]^2-\frac{g_-\rho^2 x^2}{2}.
\end{eqnarray} 
By minimizing the above express with respect to $x$, we can determine the mean-field phase diagram. 

Expanding Eq.~\eqref{Eq:MFEnergy} around $x=0$ and rewriting the result in terms of $k_1$, we get
\begin{eqnarray}\label{Eq:MFEnergyExpansion}
\mathcal{E}_{\mathrm{mf}}
&=&\frac{k^2_0}{2}\rho -\frac{\Omega}{2}\rho +\frac{g_+\rho^2}{2}+\frac{\rho}{2}\left[1-\frac{2k^2_0}{\Omega+2G_2}\right]k ^2_1.
\end{eqnarray}
It is then clear that the mean-field phase transition point is determined by $1-\frac{2k^2_0}{\Omega+2G_2}=0$. 
In the zero momentum phase, the coefficient before $k_1$ measures the energy cost of the phase fluctuations, and therefore it is by definition the superfluid density $\rho_x/2$. 
From the point view of the Landau theory of phase transitions, the superfuid density is the coefficient of the quadratic term in the order parameter expansion. 
A negative superfluid density simply means that the zero momentum phase is unstable, and $k_1$ will acquire a nonzero expectation value such that the system enters the plane wave phase. In the plane wave phase, the superfluid density becomes positive again.

To calculate the correction to the mean-field phase boundary, we include the LHY contribution to the ground state energy density
and minimize $\mathcal{E}_{\mathrm{mf}}+\mathcal{E}_{\mathrm{LHY}}$ as a function of $k_1$. 
The LHY energy is obtained through the excitation energy determined by $\det{\mathcal{G}^{-1}_0}=0$ with $\mathcal{G}^{-1}_0$ given by Eq.~\eqref{Eq:Green_PW}. 
The minimization can be done in the following way: We first calculate numerically the LHY energy for small $k_1$, and then extract the coefficient of the $k^2_1$ term in $\mathcal{E}_{\mathrm{LHY}}$. 
This coefficient gives a correction to the coefficient of $k^2_1$ in Eq.~\eqref{Eq:MFEnergyExpansion}, and the new phase boundary is determined by requiring the corrected coefficient to be zero. 
As shown in Fig.~\ref{Fig:Dxx} (d), the phase boundary determined in this way agrees perfectly with the one determined through the one-loop result of $\rho_{x}$.
  
Before closing this section, we mention that the same method can be used to get the superfluid density in the plane wave phase.  
Assuming $\mathcal{E}_{\mathrm{mf}}$ reaches its minimal at $k_{1,c}$, then the superfluid density is obtained by expanding the mean-field energy Eq.~\eqref{Eq:MFEnergy} around $k_{1,c}$,
\begin{eqnarray}
\mathcal{E}_{\mathrm{mf}}=\mathcal{E}_{\mathrm{mf}}(k_{1,c})+\frac{\rho_x}{2}\delta k^2,
\end{eqnarray}
where $\delta k=k_1-k_{1,c}$. To find $k_{1,c}$ we minimize Eq.~\eqref{Eq:MFEnergy} with respect to $x$ and find the position $x_c$ at which the energy takes minimum. Then using Eq.~\eqref{Eq:Cond2}, we find $k_{1,c}=k_0\sqrt{1-\Omega^2/\Omega^2_{c,\mathrm{mf}}}$. Expanding Eq.~\eqref{Eq:MFEnergy} around $x_c$ and change the variable from $x-x_c$ to $k_1-k_{1,c}$, we obtain
the mean-field superfluid density in the plane wave phase
\begin{eqnarray}
\rho_{x}=\rho-\rho\frac{k^2_0\Omega^2}{\Omega^2G_2 +4(k^2_0-G_2)^3},
\end{eqnarray}
which is the same as the result in \cite{PhysRevA.94.033635}. 
By taking into account the LHY contribution, we can also obtain the correction to the mean-field superfluid density in the plane wave phase.
   
\section{The damping rate at zero and finite temperature} \label{app:damping}

The damping rate $\gamma$, i.e., the imaginary part of the phonon excitation energy, is determined by
\begin{eqnarray}\label{Eq:damping}
\det [G^{-1}_0(\varepsilon_{\mathrm{ph}}-i\gamma ,\mathbf{q})-\Sigma(\varepsilon_{\mathrm{ph}}+i 0^+,\mathbf{q})] =0,
\end{eqnarray}
where $\Sigma(\varepsilon_{\mathrm{ph}}+i 0^+,\mathbf{q})$ is the one-loop self-energy evaluated at the phonon frequency.

To solve Eq.~\eqref{Eq:damping}, we first  integrate out the $\varphi$ and $\zeta_-$ fields and obtain an effective theory for the low energy mode [c.f. Eq.~\eqref{Eq:EffetiveLagrangian}]
\begin{eqnarray}
 \mathcal{L}_{\mathrm{eff}}&=&\frac{1}{2}[\phi,\zeta_+]\mathcal{G}^{-1}_{\mathrm{eff}}[\phi,\zeta_+]^{T},
\end{eqnarray}
where $\mathcal{G}^{-1}_{\mathrm{eff}}$ can be written as
\begin{eqnarray}\label{Eq:Green_eff}
 \mathcal{G}^{-1}_{\mathrm{eff}}&=&\left[\begin{array}{cc}
 \rho (\mathbf{q}^2 -\frac{2k^2_0}{\Omega +2G_2} q^2_x )  & -\omega_n  \\ 
 \omega_n & \frac{\mathbf{q}^2}{4\rho}+g_+ 
 \end{array} 
 \right] - \left[\begin{array}{cc}
   \Sigma_{\mathrm{eff},\phi\phi} & \Sigma_{\mathrm{eff},\phi\zeta_+}  \\ 
  \Sigma_{\mathrm{eff},\zeta_+\phi} & \Sigma_{\mathrm{eff},\zeta_+\zeta_+} 
  \end{array} 
  \right].
\end{eqnarray}
And then from Eq.~\eqref{Eq:Green_eff}, the damping rate is obtained
\begin{eqnarray}
\gamma=\frac{\rho (\mathbf{q}^2 -\frac{2k^2_0}{\Omega +2G_2} q^2_x )\Im \Sigma_{\mathrm{eff},\zeta_+\zeta_+} + g_+\Im \Sigma_{\mathrm{eff},\phi\phi}}{2\varepsilon_{\mathrm{ph}}} + \Re \Sigma_{\mathrm{eff},\phi\zeta_+}.
\end{eqnarray}

We focus on the linear dispersion regime defined through Eqs.~\eqref{Eq:Linear1}-\eqref{Eq:Linear3}. 
By analyzing the low energy and momentum behavior of all the one-loop self-energies, we find that it is enough to consider the Feynman diagrams constructed from only two vertices Figs.~\ref{Fig:Vertices} (a) and (d), and the momentum dependence of vertex  Fig.~\ref{Fig:Vertices} (d) can be neglected. 
Therefore the relevant parts of the effective self-energy matrix is
\begin{eqnarray}
 &&\left[\begin{array}{cc}
 \Im \Sigma_{\mathrm{eff},\phi\phi} & \Re \Sigma_{\mathrm{eff},\phi\zeta_+}  \\ 
 \Re \Sigma_{\mathrm{eff},\zeta_+\phi} & \Im \Sigma_{\mathrm{eff},\zeta_+\zeta_+} 
 \end{array} 
 \right]=\left[\begin{array}{cc}
 \Im \Sigma_{\phi\phi} & \Re \Sigma_{\phi\zeta_+}  \\ 
 \Re \Sigma_{\zeta_+\phi} & \Im \Sigma_{\zeta_+\zeta_+} 
 \end{array} 
 \right]+
 \left[\begin{array}{cc}
  -\frac{4\rho k_0 q_x \Re \Sigma_{\phi\zeta_-}}{\Omega+2G_2}+\frac{4\rho^2 k^2_0 q^2_x \Re \Sigma_{\zeta_-\zeta_-}}{(\Omega+2G_2)^2} & -\frac{2\rho k_0 q_x\Im \Sigma_{\zeta_+\zeta_-}}{\Omega+2G_2}  \\ 
  \frac{2\rho k_0 q_x\Im \Sigma_{\zeta_+\zeta_-}}{\Omega+2G_2} & 0 
  \end{array} 
  \right].
\end{eqnarray}

\begin{figure}
 	\includegraphics[width=0.5\columnwidth]{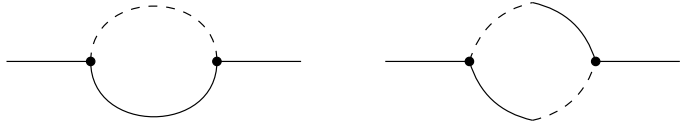}
 	\caption{Feynman diagrams for $\Sigma_{\phi\phi}$ that contribute to the phonon damping rates.}\label{Fig:Sigma_11}
\end{figure}

As an example, we calculate $\Sigma_{\phi\phi}(i\omega_n,\mathbf{q})$ explicitly. The Feynman diagrams are shown in Fig.~\ref{Fig:Sigma_11}.
\begin{eqnarray}
\Sigma_{\phi\phi}(i\omega_n,\mathbf{q})&=&\sum_{\omega'_m,\mathbf{k}}\bigg[(\mathbf{k}\cdot\mathbf{q})^2 \mathcal{G}_{0,\phi\phi}(i\omega'_m,\mathbf{k}) \mathcal{G}_{0,\zeta_+\zeta_+}(i\omega_n-i\omega'_m,\mathbf{q-k})]\nonumber\\
&&+\mathbf{k}\cdot\mathbf{q}(\mathbf{q}^2-\mathbf{k}\cdot\mathbf{q}) \mathcal{G}_{0,\phi\zeta_+}(i\omega'_m,\mathbf{k}) \mathcal{G}_{0,\phi\zeta_+}(i\omega_n-i\omega'_m,\mathbf{q-k})\bigg].
\end{eqnarray}
We write the noninteracting Green's function explicitly
\begin{eqnarray}
\mathcal{G}_{0,\phi\phi}(i\omega_n,\mathbf{q})&=&\frac{A_{11}(\mathbf{q})}{\omega^2_n+\varepsilon^2_\mathrm{sp}(\mathbf{q})}+\frac{B_{11}(\mathbf{q})}{\omega^2_n+\varepsilon^2_\mathrm{ph}(\mathbf{q})},\\
\mathcal{G}_{0,\zeta_+\zeta_+}(i\omega_n,\mathbf{q})&=&\frac{A_{22}(\mathbf{q})}{\omega^2_n+\varepsilon^2_\mathrm{sp}(\mathbf{q})}+\frac{B_{22}(\mathbf{q})}{\omega^2_n+\varepsilon^2_\mathrm{ph}(\mathbf{q})},\\ \mathcal{G}_{0,\phi\zeta_+}(i\omega_n,\mathbf{q})&=&-\mathcal{G}_{0,\zeta_+\phi}(i\omega_n,\mathbf{q})
=\frac{\omega A_{12}(\mathbf{q})}{\omega^2_n+\varepsilon^2_\mathrm{sp}(\mathbf{q})}+\frac{\omega B_{12}(\mathbf{q})}{\omega^2_n+\varepsilon^2_\mathrm{ph}(\mathbf{q})}.
\end{eqnarray}
Since we are studying the damping rate in the linear regime, the gapped branch can be neglected, and it is enough to know the low momentum behavior of $B_{11}$, $B_{22}$, and $B_{12}$,
\begin{eqnarray}
B_{11}(\mathbf{q})\approx g_+, ~ B_{22}(\mathbf{q})\approx \rho \frac{c^2_\theta}{c^2_0}\mathbf{q}^2, ~ B_{12}(\mathbf{q})\approx 1.
\end{eqnarray}
Evaluating the Matsubara frequency summation, $\Sigma_{\phi\phi}(i\omega_n,\mathbf{q})$ can be written as
\begin{eqnarray}
\Sigma_{\phi\phi}(i\omega_n,\mathbf{q})=\Sigma_{\phi\phi,1}(i\omega_n,\mathbf{q})+\Sigma_{\phi\phi,2}(i\omega_n,\mathbf{q}),
\end{eqnarray}
with
\begin{eqnarray}
\Sigma_{\phi\phi,1}(i\omega_n,\mathbf{q})&=&\sum_{\mathbf{k}}\left[1+n(\varepsilon_{\mathrm{ph}}(\mathbf{k}))+n(\varepsilon_{\mathrm{ph}}(\mathbf{q-k}))\right]
\left[\frac{1}{-i\omega_n +\varepsilon_{\mathrm{ph}}(\mathbf{k})+ \varepsilon_{\mathrm{ph}}(\mathbf{q-k}) } + \frac{1}{i\omega_n +\varepsilon_{\mathrm{ph}}(\mathbf{k})+ \varepsilon_{\mathrm{ph}}(\mathbf{q-k}) }\right]\nonumber\\
&&%
\left[(\mathbf{k}\cdot\mathbf{q})^2 \frac{B_{11}(\mathbf{k})B_{22}(\mathbf{q-k})}{4\varepsilon_{\mathrm{ph}}(\mathbf{k}) \varepsilon_{\mathrm{ph}}(\mathbf{q-k})} + \mathbf{k}\cdot\mathbf{q}(\mathbf{q}^2-\mathbf{k}\cdot\mathbf{q})\frac{B_{12}(\mathbf{k})B_{12}(\mathbf{q-k})}{4}\right],
\end{eqnarray}
which is nonzero even if the temperature is zero and is relevant to the Beliaev damping rate, and
\begin{eqnarray}
\Sigma_{\phi\phi,2}(i\omega_n,\mathbf{q})&=&\sum_{\mathbf{k}}\left[n(\varepsilon_{\mathrm{ph}}(\mathbf{k}))-n(\varepsilon_{\mathrm{ph}}(\mathbf{q-k}))\right]\left[\frac{1}{i\omega_n + \varepsilon_{\mathrm{ph}}(\mathbf{q-k})-\varepsilon_{\mathrm{ph}}(\mathbf{k}) } - \frac{1}{i\omega_n +\varepsilon_{\mathrm{ph}}(\mathbf{k})- \varepsilon_{\mathrm{ph}}(\mathbf{q-k}) }\right]\nonumber\\
&&%
\left[(\mathbf{k}\cdot\mathbf{q})^2 \frac{B_{11}(\mathbf{k})B_{22}(\mathbf{q-k})}{4\varepsilon_{\mathrm{ph}}(\mathbf{k}) \varepsilon_{\mathrm{ph}}(\mathbf{q-k})} - \mathbf{k}\cdot\mathbf{q}(\mathbf{q}^2-\mathbf{k}\cdot\mathbf{q})\frac{B_{12}(\mathbf{k})B_{12}(\mathbf{q-k})}{4}\right],
\end{eqnarray}
which is nonzero only at finite temperature and is relevant to the Landau damping rate.

We calculate the imaginary part of  $\Sigma_{\phi\phi,1}(\varepsilon_{\mathrm{ph}}(\mathbf{q})+i0^+,\mathbf{q})$ at zero temperature,
\begin{eqnarray}
\Im \Sigma_{\phi\phi,1}(\varepsilon_{\mathrm{ph}}(\mathbf{q})+i0^+,\mathbf{q})&=&\pi\sum_{\mathbf{k}}
\delta(-\varepsilon_{\mathrm{ph}}(\mathbf{q}) +\varepsilon_{\mathrm{ph}}(\mathbf{k})+ \varepsilon_{\mathrm{ph}}(\mathbf{q-k}) )f(\mathbf{q},\mathbf{k}),\\
f(\mathbf{q},\mathbf{k})&=&%
(\mathbf{k}\cdot\mathbf{q})^2 \frac{B_{11}(\mathbf{k})B_{22}(\mathbf{q-k})}{4\varepsilon_{\mathrm{ph}}(\mathbf{k}) \varepsilon_{\mathrm{ph}}(\mathbf{q-k})} + \mathbf{k}\cdot\mathbf{q}(\mathbf{q}^2-\mathbf{k}\cdot\mathbf{q})\frac{B_{12}(\mathbf{k})B_{12}(\mathbf{q-k})}{4}.
\end{eqnarray}
To calculate the above integral, we need to solve the  internal  $\mathbf{k}$ allowed by the energy and momentum conservation. We can scale the momentum as $c_x k_x \equiv c_0 k'_x$ and $k_{y/z}=k'_{y/z}$, and then the phonon dispersion can be written as
\begin{eqnarray}
\varepsilon_{\mathrm{ph}}(\mathbf{k})=\sqrt{c^2_x k^2_x+c_0 k^2_y+c^2_0 k^2_z}=c_\theta k=c_0 k'.
\end{eqnarray}
The momentum and energy conservation can be solved in terms of the new variables in the small $\mathbf{q}$ limit ($\theta'$ is the angle between $\mathbf{k}'$ and $\mathbf{q}'$),
\begin{eqnarray}
\delta(-c_0 q'+c_0 k'+ c_0 |\mathbf{q}'-\mathbf{k}'|)=\frac{q'-k'}{c_0 q' k'\sin{\theta'}}\delta(\theta'),
\end{eqnarray}
with the restrition $k'<q'$.
This means that $\mathbf{k}'$ and $\mathbf{q}'$ are along the same direction and $k'<q'$ and therefore $\mathbf{k}$ and $\mathbf{q}$ are also along the same direction and $k<q$. 
Under this condition,
\begin{eqnarray}
f(\mathbf{q},\mathbf{k})=\frac{k q^2(q-k)}{2},
\end{eqnarray}
so
\begin{eqnarray}
\Im \Sigma_{\phi\phi,1}(\varepsilon_{\mathrm{ph}}(\mathbf{q})+i0^+,\mathbf{q})&=&\pi\int\frac{\mathrm{d}k_x \mathrm{d}k_y \mathrm{d}k_z}{(2\pi)^3}f(\mathbf{q},\mathbf{k}) \delta(-\varepsilon_{\mathrm{ph}}(\mathbf{q}) +\varepsilon_{\mathrm{ph}}(\mathbf{k})+ \varepsilon_{\mathrm{ph}}(\mathbf{q-k}) ),\\
&=&\pi\frac{c_0}{c_x}\int\frac{\mathrm{d}k'_x \mathrm{d}k'_y \mathrm{d}k'_z}{(2\pi)^3}f(\mathbf{q},\mathbf{k})\delta(-c_0 q'+c_0 k'+ c_0 |\mathbf{q}'-\mathbf{k}'|),\\
&=&\pi\frac{c_0}{c_x}\int\frac{\mathrm{d}k'_x \mathrm{d}k'_y \mathrm{d}k'_z}{(2\pi)^3}f(q,k)\frac{q'-k'}{c_0 q'k' \sin{\theta'}}\delta(\theta'),\\
&=&
\pi
\frac{1}{c_x}\int\frac{k'^2\mathrm{d}k'}{4\pi^2}f(q,k)\frac{q'-k'}{q'k' },\\
&=&\frac{c^2_\theta}{c_0^2c_x}
\int^q_0\frac{k^2\mathrm{d}k}{4\pi}\frac{k q^2(q-k)}{2}\frac{q-k}{qk },\\
&=&\frac{1}{2}\frac{c^2_\theta}{c_0^2 c_x}\frac{q^6}{120\pi}.
\end{eqnarray}
We now calculate $\Im\Sigma_{\phi\phi,2}(\varepsilon_{\mathrm{ph}}(\mathbf{q})+i0^+,\mathbf{q})$ at finite temperature,
\begin{eqnarray}
\Im\Sigma_{\phi\phi,2}(\varepsilon_{\mathrm{ph}}(\mathbf{q})+i0^+,\mathbf{q})&=&\pi\sum_{\mathbf{k}}\left[n(\varepsilon_{\mathrm{ph}}(\mathbf{k}))-n(\varepsilon_{\mathrm{ph}}(\mathbf{q-k}))\right]\left[(\mathbf{k}\cdot\mathbf{q})^2 \frac{B_{11}(\mathbf{k})B_{22}(\mathbf{q-k})}{4\varepsilon_{\mathrm{ph}}(\mathbf{k}) \varepsilon_{\mathrm{ph}}(\mathbf{q-k})} - \mathbf{k}\cdot\mathbf{q}(\mathbf{q}^2-\mathbf{k}\cdot\mathbf{q})\frac{B_{12}(\mathbf{k})B_{12}(\mathbf{q-k})}{4}\right]\nonumber\\
&&%
\left[-\delta(\varepsilon_{\mathrm{ph}}(\mathbf{q}) + \varepsilon_{\mathrm{ph}}(\mathbf{q-k})-\varepsilon_{\mathrm{ph}}(\mathbf{k})) + \delta(\varepsilon_{\mathrm{ph}}(\mathbf{q}) +\varepsilon_{\mathrm{ph}}(\mathbf{k})- \varepsilon_{\mathrm{ph}}(\mathbf{q-k}) )\right],\\
&=&-\pi\sum_{\mathbf{k}}\left[n(\varepsilon_{\mathrm{ph}}(\mathbf{k+q}))-n(\varepsilon_{\mathrm{ph}}(\mathbf{k}))\right]\delta(\varepsilon_{\mathrm{ph}}(\mathbf{q}) + \varepsilon_{\mathrm{ph}}(\mathbf{k})-\varepsilon_{\mathrm{ph}}(\mathbf{k+q}))g(\mathbf{q},\mathbf{k}),\\
&=&-\pi\sum_{\mathbf{k}}\frac{\partial n(\varepsilon_{\mathrm{ph}}(\mathbf{k}))}{\partial \varepsilon_{\mathrm{ph}}(\mathbf{k})}\varepsilon_{\mathrm{ph}}(\mathbf{q})\delta(\varepsilon_{\mathrm{ph}}(\mathbf{q}) + \varepsilon_{\mathrm{ph}}(\mathbf{k})-\varepsilon_{\mathrm{ph}}(\mathbf{k+q}))g(\mathbf{q},\mathbf{k}),\label{Eq:ImSigma}
\end{eqnarray}
where
\begin{eqnarray}
g(\mathbf{q},\mathbf{k})&=&\left[(\mathbf{k}\cdot\mathbf{q}+\mathbf{q}^2)^2 \frac{B_{11}(\mathbf{k+q})B_{22}(\mathbf{k})}{4\varepsilon_{\mathrm{ph}}(\mathbf{k+q}) \varepsilon_{\mathrm{ph}}(\mathbf{k})} + (\mathbf{k}\cdot\mathbf{q}+\mathbf{q}^2)\mathbf{k}\cdot\mathbf{q}\frac{B_{12}(\mathbf{k+q})B_{12}(\mathbf{k})}{2}+(\mathbf{k}\cdot\mathbf{q})^2 \frac{B_{11}(\mathbf{k})B_{22}(\mathbf{q+k})}{4\varepsilon_{\mathrm{ph}}(\mathbf{k}) \varepsilon_{\mathrm{ph}}(\mathbf{q+k})}\right].~~~~~~
\end{eqnarray}
To get Eq.~\eqref{Eq:ImSigma} we  have assumed  $c_\theta q/T\ll 1$ and expand $n(\varepsilon_{\mathrm{ph}}(\mathbf{k+q}))-n(\varepsilon_{\mathrm{ph}}(\mathbf{k}))$ to the lowest order. 
In general it is difficult to solve the energy and momentum conserving condition $\delta(\varepsilon_{\mathrm{ph}}(\mathbf{q}) + \varepsilon_{\mathrm{ph}}(\mathbf{k})-\varepsilon_{\mathrm{ph}}(\mathbf{k+q}))$ even if $\mathbf{q}$ is small, because  $\mathbf{k}$ is not necessarily small and for general $\mathbf{k}$, the phonon dispersion is very complicated. 
However, if we focus on the low temperature region such that the corresponding phonon dispersion is linear, then we can replace  $\varepsilon_{\mathrm{ph}}(\mathbf{k})$ by the linear dispersion because  $\frac{\partial n(\varepsilon_{\mathrm{ph}}(\mathbf{k}))}{\partial \varepsilon_{\mathrm{ph}}(\mathbf{k})}$ decays rapidly when $\varepsilon_{\mathrm{ph}}(\mathbf{k})>T$. 
In this region the momentum and energy conservation is easily solved: $\mathbf{k}$ and $\mathbf{q}$ are along the same direction and the length of $\mathbf{k}$ is unrestricted. Under this condition $g(\mathbf{q},\mathbf{k})$ also takes a simple form
\begin{eqnarray}
g(\mathbf{q},\mathbf{k})&=&k q^2 (k+q),
\end{eqnarray}
and  
\begin{eqnarray}
\Im\Sigma_{\phi\phi,2}(\varepsilon_{\mathrm{ph}}(\mathbf{q})+i0^+,\mathbf{q})&=&\frac{c^2_\theta}{c_0^2c_x}\int^\infty_0\frac{k^2\mathrm{d}k}{4\pi}\frac{\beta e^{\beta c_\theta k}}{(e^{\beta c_{\theta} k}-1)^2}c_{\theta} q
k q^2(k+ q)\frac{q+k}{qk },\\
&=&\frac{q^2 T^4}{ c^2_\theta c^2_0 c_x} 
\int^\infty_0\frac{x^2\mathrm{d}x}{4\pi}\frac{ e^{x}}{(e^{x}-1)^2}
\left(x+ \frac{c_\theta q}{T}\right )^2,\label{Eq:Sigma112_1}\\
&=&\frac{\pi^3 q^2 T^4}{15 c^2_\theta c^2_0 c_x}.\label{Eq:Sigma112_2}
\end{eqnarray}
To get Eq.~\eqref{Eq:Sigma112_2} from Eq.~\eqref{Eq:Sigma112_1}, we have used  the condition $c_\theta q/T\ll 1$. 

We can calculate other self-energies in the similar way, and here we just summarize the final results,
\begin{eqnarray}
\begin{array}{l|l}
\Im \Sigma_{\phi\phi,1}=\frac{1}{2}\frac{c^2_\theta}{c^2_0 c_x}\frac{q^6}{120\pi} & \Im \Sigma_{\phi\phi,2}=\frac{\pi^3 q^2 T^4}{15 c^2_\theta c^2_0 c_x} \\[5pt] 
\Re \Sigma_{\phi\zeta_+,1}=\frac{g_+}{4}\left[1-\frac{2k^2_0\Omega\cos^2{\theta}}{(\Omega+2G_2)^2} \right]\frac{c_\theta}{c^2_0 c_x}\frac{q^5}{120\pi} & \Re \Sigma_{\phi\zeta_+,2}=\frac{g_+}{2}\left[1-\frac{2k^2_0\Omega\cos^2{\theta}}{(\Omega+2G_2)^2} \right]\frac{1}{c^2_0 c^3_\theta c_x}\frac{\pi^3 q T^4}{15} \\[5pt]
\Im \Sigma_{\zeta_+\zeta_+,1}=\frac{g_+}{8\rho}\left[1-\frac{2\Omega k^ 2_0\cos^2{\theta}}{(\Omega+2G_2)^2}\right]^2 \frac{1}{ c_x}\frac{q^4}{120\pi}  & \Im \Sigma_{\zeta_+\zeta_+,2}=\frac{g_+}{4\rho}\left[1-\frac{2\Omega k^ 2_0\cos^2{\theta}}{(\Omega+2G_2)^2}\right]^2 \frac{1}{c^4_\theta c_x}\frac{\pi^3T^4}{15}  \\[5pt]
\Im \Sigma_{\zeta_-\zeta_-,1}= \frac{k^2_0\Omega^2 \cos^2{\theta}}{2\rho^2 (\Omega+2G_2)^2}\frac{c^2_\theta}{c^2_0 c_x} \frac{q^4}{120\pi} &  \Im \Sigma_{\zeta_-\zeta_-,2}= \frac{k^2_0\Omega^2 \cos^2{\theta}}{\rho^2 (\Omega+2G_2)^2}\frac{1}{c^2_0 c^2_\theta c_x} \frac{\pi^3 T^4}{15}\\[5pt] 
 \Re \Sigma_{\phi \zeta_-,1}= \frac{\Omega k_0\cos{\theta}}{2\rho (\Omega+2G_2)} \frac{c^2_\theta}{c^2_0 c_x} \frac{q^5}{120\pi}& \Re \Sigma_{\phi \zeta_-,2}= \frac{\Omega k_0\cos{\theta}}{\rho (\Omega+2G_2)} \frac{1}{c^2_0 c^2_\theta c_x} \frac{\pi^3 q T^4}{15}\\[5pt]
 \Im \Sigma_{\zeta_+ \zeta_-,1}= \frac{g_+ \Omega k_0\cos{\theta}}{4\rho (\Omega+2G_2)}
 \left[1-\frac{2k^2_0\Omega\cos^2{\theta}}{(\Omega+2G_2)^2} \right]
 \frac{c_\theta}{c^2_0 c_x} \frac{q^4}{120\pi} & \Im \Sigma_{\zeta_+ \zeta_-,2}= \frac{g_+ \Omega k_0\cos{\theta}}{2\rho (\Omega+2G_2)}
  \left[1-\frac{2k^2_0\Omega\cos^2{\theta}}{(\Omega+2G_2)^2} \right]
  \frac{1}{c^2_0 c^3_\theta c_x} \frac{\pi^3 T^4}{15}
\end{array} 
\end{eqnarray}
From the above results we get the Beliaev damping rate at zero temperature
\begin{eqnarray}
\gamma_\mathrm{B}&=&\frac{3g_+q^5}{640\pi c_\theta}\left[1-\frac{2\Omega k^ 2_0\cos^2{\theta}}{(\Omega+2G_2)^2}\right]^2 \frac{c^2_\theta}{c^2_0c_x},\\
&=&\frac{3q^5}{640\pi\rho}\left[1-\frac{2\Omega k^ 2_0\cos^2{\theta}}{(\Omega+2G_2)^2}\right]^2 \frac{c_\theta}{c_x},\\
&=&\frac{3q^5}{640\pi\rho}\left[1-\frac{2\Omega k^ 2_0\cos^2{\theta}}{(\Omega+2G_2)^2}\right]^2 \sqrt{1+\frac{2k^2_0\sin^2{\theta}}{\Omega+2G_2-2k^2_0}}.\nonumber\\
\end{eqnarray}
and the Landau damping rate at finite temperature
\begin{eqnarray}
\gamma_\mathrm{L}=\frac{3\pi^3 q T^4}{40\rho c^4_\theta}\left[1-\frac{2\Omega k^ 2_0\cos^2{\theta}}{(\Omega+2G_2)^2}\right]^2 \sqrt{1+\frac{2k^2_0\sin^2{\theta}}{\Omega+2G_2-2k^2_0}}.\nonumber\\
\end{eqnarray}
The Beliaev damping rate takes the same form as the result in \cite{Beliaev_SOC}, where a different method was used. 
The analytical expression for the Landau damping rate is obtained here for the first time.

If $G_2=g_-\rho=0$, the damping rates can be further simplified as
\begin{eqnarray}
\gamma_\mathrm{B}&=&\frac{3q^5}{640\pi\rho}\frac{c^4_\theta}{c^4_0} \sqrt{1+\frac{2k^2_0\sin^2{\theta}}{\Omega-2k^2_0}},\\
\gamma_\mathrm{L}&=&\frac{3\pi^3 q T^4}{40\rho c^4_0} \sqrt{1+\frac{2k^2_0\sin^2{\theta}}{\Omega-2k^2_0}}.
\end{eqnarray}
Since $c_\theta=c_0\sqrt{1-\frac{2k^2_0\cos^2{\theta}}{\Omega+2G_2}}$, the Beliaev damping is strongly suppressed when the momentum is along the direction of the spin-orbit coupling. 
However, the Landau damping is not suppressed. 
\end{widetext}

\bibliography{soc_boson.bib}

\end{document}